\PassOptionsToPackage{hypertexnames=false}{hyperref}
\documentclass[journal]{IEEEtran}

\usepackage[T1]{fontenc}
\usepackage{amsmath,amssymb,amsthm,mathtools}
\usepackage{bm}
\usepackage{booktabs}
\usepackage{multirow}
\usepackage{array}
\usepackage{graphicx}
\usepackage{xcolor}
\usepackage{colortbl}
\usepackage{cite}
\usepackage{url}
\usepackage{orcidlink}
\usepackage{hyperref}
\hypersetup{colorlinks=true, urlcolor=blue, linkcolor=black, citecolor=blue}

\newtheorem{theorem}{Theorem}

\newtheorem{corollary}[theorem]{Corollary}
\newtheorem{definition}{Definition}

\usepackage{algorithm}
\usepackage{float}
\usepackage{enumitem}  
\makeatletter
\renewcommand{\fs@ruled}{\def\@fs@cfont{\bfseries}\let\@fs@capt\floatc@ruled
  \def\@fs@pre{\hrule height.8pt depth0pt \kern2pt}%
  \def\@fs@post{\kern2pt\hrule height.8pt\relax}%
  \def\@fs@mid{\kern2pt\hrule height.4pt\kern2pt}%
  \let\@fs@iftopcapt\iftrue}
\makeatother
\floatstyle{ruled}
\restylefloat{algorithm}

\newcommand{\R}{\mathbb{R}}

\newcommand{\D}{\mathcal{D}}

\renewcommand{\arraystretch}{1.08}
\title{Decision-Aware Quadratic ReLU Replacement for HE-Friendly Inference}

\author{Rui Li\textsuperscript{\orcidlink{0009-0009-2936-6472}}, Wenyuan Wu\textsuperscript{\orcidlink{0000-0003-1169-379X}}, and Weijie Miao\textsuperscript{\orcidlink{0009-0009-9205-6207}}%
\thanks{This work was supported in part by the National Key Research Project of China under Grant 2025YFA1017201, and in part by the National Natural Science Foundation of China (NSFC) under Grant 12571553. (Corresponding author: Wenyuan Wu.)}%
\thanks{Rui Li and Wenyuan Wu are with the Chongqing Key Laboratory of Secure Computing for Biology, Chongqing Institute of Green and Intelligent Technology, Chinese Academy of Sciences, 266 Fangzheng Avenue, Beibei District, Chongqing 400714, China (e-mail: lirui232@mails.ucas.ac.cn; wuwenyuan@cigit.ac.cn).}%
\thanks{Weijie Miao is with the Department of Industrial and Systems Engineering, The Hong Kong Polytechnic University, Hung Hom, Hong Kong, China (e-mail: weijie.miao@connect.polyu.hk).}}

\begin{document}

\markboth{IEEE Transactions on Information Forensics and Security}{Li \MakeLowercase{\textit{et al.}}: Decision-Aware Quadratic ReLU Replacement for HE-Friendly Inference}

\maketitle

\begin{abstract}
Fully homomorphic encryption (FHE) supports only additions and
multiplications, so FHE-only neural-network inference typically replaces
ReLU with polynomials fitted over empirical activation intervals. Such
interval fitting often requires higher-degree polynomials to control
activation error, incurring homomorphic evaluation costs, while classification is determined by the final logit decision. We revisit ReLU
replacement from a decision-aware perspective: given a trained
single-hidden-layer ReLU MLP and a specified calibration set, can an
HE-friendly low-degree polynomial replace ReLU without retraining while
preserving calibration-set decisions? We focus on quadratic replacement,
the lowest-degree that retains a genuine per-unit nonlinearity.
For calibration sets positive-margin separable in the
lifted space, we formulate quadratic replacement as a linear separation
problem, yielding necessary and sufficient conditions for
calibration-lossless replacement and a constructive algorithm for the
coefficients. When the positive-margin condition fails---often because a few near-boundary or misclassified calibration samples bring the lifted hulls into contact---we extend the same
geometric framework via reduced convex hulls and Lagrangian-dual
soft-margin relaxations. These cap the weight any single sample can carry,
converting the problem into smaller convex quadratic programs that yield
approximately feasible coefficients with high empirical agreement on
calibration-set decisions. In particular, at the maximal weight cap
$\mu=1$, the reduced-convex-hull relaxation reduces to standard convex-hull
separation; the relaxation thus continuously extends the positive-margin
exact theory. Under CKKS, the quadratic
replacement matches plaintext top-1 accuracy on multiple benchmarks,
running 3.7--4.1$\times$ faster than Remez-7 in the activation module and
1.18--1.68$\times$ faster end-to-end.
\end{abstract}


\section{Introduction}

\IEEEPARstart{F}{ully} homomorphic encryption (FHE)
enables a server to evaluate arithmetic circuits over encrypted data without
decrypting the input~\cite{rivest1978privacy,gentry2009fhe,cheon2017ckks}.
This property makes FHE a natural tool for privacy-preserving neural-network
inference. However, encrypted computation is efficient mainly when the network
can be expressed as a low-depth circuit of additions and multiplications.
Affine layers fit this model, whereas comparisons, max operations, and
ReLU activations are costly when inference uses only
FHE arithmetic. We use \emph{FHE-only inference} to mean this noninteractive
arithmetic-circuit route, without interactive or hybrid nonlinear subprotocols. This tension makes nonlinear activations a central
issue in HE-friendly neural-network inference~\cite{giladbachrach2016cryptonets,
hesamifard2017cryptodl,brutzkus2019lola,dathathri2019chet,dathathri2020eva,
lee2023precise,lee2023ola,ao2024autofhe}.

Existing private-inference systems address nonlinear activations through
several routes. The first route is FHE-only evaluation: the network is
compiled into a polynomial arithmetic circuit, and nonlinear activations are
replaced by polynomial functions~\cite{giladbachrach2016cryptonets,
hesamifard2017cryptodl,brutzkus2019lola,dathathri2019chet,dathathri2020eva,
boemer2019ngraph}. This route keeps the online protocol simple, but its
accuracy and efficiency depend strongly on polynomial degree, multiplicative
depth, rescaling, and modulus-chain budget. The second route is secure
multiparty computation (MPC), which can evaluate comparisons and ReLU-like
operations through interactive protocols, but pays communication,
synchronization, and system-level costs~\cite{mohassel2017secureml,
liu2017minionn,riazi2019xonn,rathee2020cryptflow2}. The third route combines
FHE with MPC or garbled circuits, using FHE for linear layers and interactive
protocols for nonlinear layers~\cite{juvekar2018gazelle,mishra2020delphi,
huang2022cheetah}. These hybrid systems can support exact or high-fidelity
nonlinearities, but they also introduce protocol complexity. This paper
follows the FHE-only route and asks how far one can go with a very low-degree,
HE-friendly polynomial replacement.

A common FHE-only strategy is to replace ReLU by a polynomial fitted over an
empirical activation interval. The polynomial may be the square function, a
Chebyshev or least-squares approximation, a Remez/minimax polynomial, or a
composite polynomial approximation~\cite{trefethen2013,hesamifard2017cryptodl,
lee2023precise,lee2023ola,lou2021safenet,ao2024autofhe}. Such methods are
FHE-compatible because they avoid comparisons, but high-degree polynomials are
not necessarily HE-friendly: controlling activation error over a broad interval
often requires higher degree, incurring additional ciphertext-ciphertext
multiplications, rescaling operations, modulus-chain length, and latency. Their
design objective is \emph{module-local}, meaning that the activation module is
optimized in isolation. A typical replacement solves a scalar approximation
problem of the form
\begin{equation}
\min_{p\in\Pi_d}\;\sup_{u\in[a,b]}
\left|p(u)-\max\{0,u\}\right|,
\label{eq:scalar-approx}
\end{equation}
where $[a,b]$ is chosen from activation statistics or from a conservative
range.

Objective~\eqref{eq:scalar-approx} is a meaningful numerical approximation
criterion, but it is only a proxy for the deployed classification objective.
A classifier returns the label with the largest final output logit; hence the
relevant event is whether the final top-1 logit ordering is preserved. Equivalently, a replacement is
harmless for a sample as long as its score perturbation does not cross the
sample's top-1/top-2 margin; the same activation error can be harmless for a
large-margin sample and harmful for a low-margin sample. This margin view is
consistent with the geometry of margin-based
classification~\cite{cortes1995svm,bennett2000svm} and with the broader
observation that small perturbations can change neural-network decisions when
they align with vulnerable directions~\cite{goodfellow2015adversarial}. Conversely, a polynomial that is
not a close pointwise approximation to ReLU at intermediate activations may
still preserve all final top-1 decisions on a declared calibration set.
Fixed-interval approximations also inherit a range-selection issue: if future
pre-activations fall outside the interval used for fitting, polynomial behavior
is no longer controlled by the stated approximation error, a classical concern
in polynomial approximation~\cite{trefethen2013}.

To obtain a mathematically transparent yet practically common setting, this
paper studies single-hidden-layer ReLU MLPs and MLP
heads on frozen representations. A single replaced nonlinear layer is the
minimal setting in which the trained hidden representation, the output head,
and the final decision rule can be analyzed exactly. The setting covers tabular
MLPs as well as DINOv2 visual features and Qwen3-Embedding text features
followed by MLP heads~\cite{oquab2023dinov2,qwen2025qwen3embedding}. Related
representation-transfer and student-model settings also use compact MLP or
MLP-like heads/backbones~\cite{tolstikhin2021mlpmixer,touvron2023resmlp,
zhang2022glnn}. For frozen-backbone experiments, the privacy claim is encrypted
MLP-head inference: the feature extractor runs on the client and the resulting
embedding is encrypted before server-side head evaluation.

\paragraph*{Theoretical formulation.}
Instead of first approximating ReLU locally and then hoping the final
classifier remains unchanged, we globally impose conditions on the final logit
ordering induced by the trained classifier. Given a trained single-hidden-layer
ReLU MLP and a specified calibration set, we ask whether ReLU can be replaced by
a low-degree polynomial, without retraining, while preserving the model's
decisions on that calibration set. The coefficients are computed offline; the
online encrypted circuit contains only the resulting polynomialized model.

We focus on a shared quadratic replacement
\begin{equation}
q(u;\alpha,\beta,\eta)=\alpha u^2+\beta u+\eta.
\label{eq:quad-form}
\end{equation}
This is the lowest-degree replacement that retains a genuine per-unit nonlinear
response: the shared quadratic acts on each hidden pre-activation $y_j$ through
the self-term $y_j^2$, introducing no cross-unit terms $y_jy_k$. A linear
replacement would instead collapse a single-hidden-layer ReLU MLP into an affine
classifier once the trained head is fixed, whereas degrees above two require
additional encrypted multiplications and rescalings. Quadratic replacement is
therefore a natural boundary point: it keeps nonlinearity while remaining
HE-friendly.

Our exact theory starts from positive-margin separability in the lifted
calibration space. For binary classification, quadratic replacement reduces to
positive-margin hyperplane separation of two lifted convex
hulls~\cite{boyd2004convex}: this condition is necessary and sufficient for
calibration-lossless replacement, and when it holds valid quadratic coefficients
can be constructed. For multiclass heads, the corresponding condition is strict
feasibility of pairwise affine logit-margin inequalities in the quadratic
coefficients.

Many practical calibration sets fail these positive-margin conditions under such
a small quadratic family. When positive-margin separation or pairwise margin
feasibility is unavailable, we extend the same geometric framework through
reduced-convex-hull and Lagrangian-dual soft-margin
relaxations~\cite{cortes1995svm,bennett2000svm}, which convert the calibration
logit-ordering constraints into small convex quadratic programs. The relaxed
solution is not an exact certificate; it is an approximately feasible set of
coefficients, reported together with margins, slacks, and calibration/test
agreement diagnostics.

\paragraph*{Contributions.}
The contributions of this paper are as follows.

\begin{itemize}
\item \emph{Decision-aware formulation for HE-friendly ReLU replacement.}
We formulate post-training activation replacement as a calibration-set
logit-ordering problem rather than a module-local ReLU approximation problem.
The benefit is direct alignment with classification output: the polynomial is
chosen to preserve the trained model's top-1 decisions on the declared
calibration set, while the online encrypted circuit remains a low-degree
arithmetic circuit.

\item \emph{Exact finite-sample theory via positive-margin separation.}
For trained single-hidden-layer ReLU MLPs, we use algebraic identity
transformations to express quadratic replacement through lifted statistics
induced by the fixed model. In the binary case, exact calibration-lossless
replacement is possible if and only if the lifted positive and negative convex
hulls admit a positive-margin hyperplane separator; when feasible, the
closest-pair construction yields quadratic coefficients and a margin
diagnostic in $O(n\log n)$ time after the lifted statistics are computed. In the
multiclass case, exact calibration-lossless preservation becomes strict
feasibility of pairwise affine logit-margin inequalities.

\item \emph{Reduced-hull and soft-margin relaxations beyond positive-margin feasibility.}
When the positive-margin conditions fail, we connect the exact geometry to
practical surrogates through reduced convex hulls and Lagrangian-dual soft-margin
relaxations. These relaxations produce smaller convex quadratic programs and
return approximately feasible coefficients with explicit slack and agreement
diagnostics, rather than claiming exact preservation where none is certified.
At the maximal weight cap, the reduced convex hull reduces to the standard
convex hull, so this relaxation continuously extends the exact theory of the
previous contribution rather than forming a separate method.

\item \emph{Algorithms, library, and encrypted validation.}
We implement the coefficient-construction pipeline in the open-source
\textsc{Quad4FHE} library,\footnote{\url{https://github.com/LeeRay629/Rethinking_ReLU_Replacement_in_MLP_Classifiers}}
including hidden-statistic computation,
hard-case testing, reduced-convex-hull search, soft-margin fallback, and
regime reporting. The coefficient search is performed offline and does not
alter the online encrypted circuit. Using the resulting coefficients, CKKS
evaluation of the quadratic replacement model matches the corresponding
plaintext top-1 accuracy on MLP/Otto, DINOv2/FGVC-Aircraft, and
Qwen3/MASSIVE, while running 3.7--4.1$\times$ faster than Remez-7 in the
activation module and 1.18--1.68$\times$ faster end-to-end.
\end{itemize}

\section{Related Work}

\subsection{HE-friendly and secure neural inference}

Early encrypted and secure machine-learning systems studied private
classification and neural-network prediction under cryptographic
constraints~\cite{bost2015ml,mohassel2017secureml,liu2017minionn}.
CryptoNets~\cite{giladbachrach2016cryptonets} showed that neural-network
inference can be evaluated over encrypted data when the network is expressed
as a low-degree arithmetic circuit. Subsequent FHE-only and HE-compiler
systems improved encrypted linear algebra, batching, packing, and circuit
optimization: LoLa optimized low-latency encrypted inference~\cite{brutzkus2019lola};
CHET~\cite{dathathri2019chet} and EVA~\cite{dathathri2020eva} compiled
homomorphic tensor computations; and nGraph-HE2~\cite{boemer2019ngraph}
integrated CKKS-based encrypted inference into a deep-learning framework.
These systems establish a common design principle for FHE-only inference:
reducing multiplicative depth and avoiding expensive nonlinear encrypted
operations are critical for practical latency.

Another line of work handles nonlinearities through MPC, garbled circuits, or
hybrid FHE--MPC protocols. GAZELLE~\cite{juvekar2018gazelle} combines
homomorphic linear layers with garbled-circuit nonlinearities. DELPHI~\cite{mishra2020delphi}
provides cryptographic inference with exact ReLU subprotocols or polynomial
approximations. CrypTFlow2, XONN, and Cheetah improve secure two-party
inference through optimized comparison, binary-network, or nonlinear-layer
protocols~\cite{rathee2020cryptflow2,riazi2019xonn,huang2022cheetah}. These
systems are complementary to ours: they securely compute nonlinearities through
additional protocol machinery, whereas we remain FHE-only and ask whether an
HE-friendly polynomial replacement can preserve the trained classifier's top-1
decisions on a declared calibration set.

\subsection{Post-training polynomial treatment of ReLU}

Polynomial activation replacement is a standard strategy for FHE-only neural-network
inference because it removes comparisons and piecewise-linear operations from
the encrypted circuit. The square activation is the cheapest nonlinear choice,
but directly substituting it for ReLU can substantially shift a trained
model's decision boundary. Classical approximation theory provides
Chebyshev, least-squares, minimax, and Remez tools for reducing pointwise
approximation error on a fixed interval~\cite{trefethen2013}. CryptoDL-style
networks adopt this polynomial viewpoint for encrypted inference~\cite{hesamifard2017cryptodl}.
More accurate post-training methods include Precise Approximation, which
composes low-degree minimax polynomials to approximate ReLU and max-pooling
without retraining~\cite{lee2023precise}; OLA, which selects layerwise
approximation degrees and modulus chains according to layer impact and input
distributions~\cite{lee2023ola}; SAFENet, which uses mixed-degree
approximations to balance accuracy and cost~\cite{lou2021safenet}; and
AutoFHE, which searches over polynomial activations and FHE evaluation
architectures~\cite{ao2024autofhe}. These approaches are FHE-compatible, but
higher-degree polynomials can still be costly in practice because they require
more encrypted multiplications, rescalings, and modulus levels.

Most post-training polynomial replacements optimize a module-local objective:
they choose an empirical activation interval and approximate ReLU, or an
activation-related surrogate, accurately on that interval. Our work is also
post-training, but the mathematical object is different. We do not fit ReLU locally on
$[a,b]$; instead, we globally fit the logit-ordering inequalities induced by the fixed
hidden layer, the fixed output head, and the declared calibration set. Thus
intermediate activation values need not be close to ReLU as long as the final
top-1 label is preserved. This decision-aware formulation turns scalar
activation approximation into a convex-geometric feasibility and relaxation
problem: positive-margin separability of the lifted calibration sets gives a
calibration-lossless certificate, while failures of the positive-margin condition
are handled by reduced-convex-hull and soft-margin convex programs.

\section{Problem Formulation}

\subsection{A single replaced layer}

Let $\D_{\mathrm{cal}}=\{x_i\}_{i=1}^n$ be a declared finite calibration set.
Before fitting the replacement polynomial, each sample is assigned a fixed
target decision $t_i$. In this paper and in all reported experiments, $t_i$ is
the decision of the original trained ReLU model on $x_i$ (the positive/negative
decision for binary classifiers, the top-1 class for multiclass classifiers).
The fitting objective is thus post-training agreement with the trained
classifier, not retraining against new labels; the same algebra would apply to
externally supplied labels, but that is not the setting evaluated here.

\begin{definition}[Calibration-lossless replacement]
\label{def:cal-lossless}
For a fixed calibration set and fixed targets $\{t_i\}_{i=1}^n$, a polynomial
replacement is \emph{calibration-lossless} if the predicted decision of the
polynomialized model equals $t_i$ for every $x_i\in\D_{\mathrm{cal}}$.
\end{definition}

\paragraph*{Scope and guarantee.}
The guarantees in this paper are deliberately scoped, and we state that scope
once here. (i) \emph{Architecture.} The construction applies to a single
replaced ReLU layer---a single-hidden-layer MLP, or an MLP head on a frozen
backbone---because the lifted geometry is planar only when one nonlinear layer
lies between fixed affine maps. (ii) \emph{Calibration dependence.} The fitted
coefficients depend on the declared calibration set, exactly as fixed-interval
baselines depend on their empirical fitting interval $[a,b]$; the trained
network is itself already data-dependent. (iii) \emph{Nature of the
certificate.} An exact certificate states only that the declared
\emph{calibration-set} decisions are unchanged; it is a finite-sample
statement, not a distribution-free guarantee for unseen inputs. A relaxed RCH
or soft-margin solution provides high empirical agreement rather than certified
losslessness, and is always reported with explicit margin, slack, and
calibration/test agreement diagnostics. (iv) \emph{Empirical claims.}
Small-pool results are single fixed-split ablations that diagnose
calibration-pool size, not estimates of sampling variance over random pools.
Generalization is assessed, as for any learned model, through held-out test
results.

We begin with a trained binary single-hidden-layer ReLU MLP
\begin{equation}
F(x)=\sum_{j=1}^{m} a_j\,\sigma\bigl(y_j(x)\bigr)+b,\qquad \sigma(u)=\max\{0,u\},
\label{eq:F-binary}
\end{equation}
where $a_j$ and $b$ are the trained output weights and bias, and $y_j(x)$
is the $j$th hidden pre-activation. The first-layer weights, output
weights, and bias are fixed; no model retraining is performed. Only the
three replacement coefficients are fitted in a post-training calibration step.

We separate the replacement process into an \emph{offline calibration phase},
in which the trained ReLU model is evaluated on the calibration set and the
coefficients $(\alpha,\beta,\eta)$ are computed, and an \emph{online encrypted
inference phase}, in which the server evaluates only the resulting
polynomialized arithmetic circuit on encrypted inputs. This separation is
essential: the coefficient computation may use plaintext calibration data,
convex-hull operations, or small convex programs, but none of these appears
inside the encrypted inference circuit.

We replace every ReLU in the layer by a shared quadratic polynomial
$q(u;\alpha,\beta,\eta)=\alpha u^2+\beta u+\eta$, giving
\begin{align}
\widetilde{F}_{\alpha,\beta,\eta}(x)
&=\sum_{j=1}^{m} a_j\bigl(\alpha y_j(x)^2+\beta y_j(x)+\eta\bigr)+b
\nonumber\\
&=\alpha\underbrace{\sum_{j} a_j\,y_j(x)^2}_{\displaystyle Q(x)}
+\beta\underbrace{\Bigl(\sum_{j} a_j\,y_j(x)+b\Bigr)}_{\displaystyle H(x)}
\nonumber\\
&\quad+\underbrace{(1-\beta)b+\eta\!\sum_{j} a_j}_{\displaystyle C_{\beta,\eta}}.
\label{eq:F-tilde-decomp}
\end{align}
Let $B=\sum_j a_j$, so that $C_{\beta,\eta}=(1-\beta)b+\eta B$.

\begin{definition}[Binary quadratic lift]
\label{def:binary-lift}
For the fixed trained binary head and the shared quadratic replacement above,
define
\begin{equation}
Q(x)=\sum_j a_j\,y_j(x)^2,\qquad H(x)=\sum_j a_j\,y_j(x)+b,
\label{eq:QH}
\end{equation}
and
\begin{equation}
\Phi(x)=\bigl(Q(x),\,H(x)\bigr)^{\!\top}\in\R^2,
\qquad
\theta=(\alpha,\beta)^{\!\top}.
\label{eq:lift}
\end{equation}
We call $\Phi(x)$ the \emph{binary quadratic lift}. It is two-dimensional
because, after the trained head is fixed, a shared quadratic replacement can
change the binary score only through $\theta^{\!\top}\Phi(x)$ plus the
sample-independent offset $C_{\beta,\eta}$.
\end{definition}

Let $\D^+$ and $\D^-$ denote the positive and negative target calibration
sets induced by the original ReLU decisions. For binary classifiers, the exact feasibility analysis below uses an adjustable post-replacement threshold. 
A replacement is calibration-lossless for the binary split if there exists a
threshold $\tau$ such that
\begin{equation}
\min_{x\in\D^+}\widetilde{F}(x)\,>\,\tau\,>\,\max_{x\in\D^-}\widetilde{F}(x).
\label{eq:thresh}
\end{equation}
Equivalently, all replaced positive scores exceed all replaced negative
scores. Since $C_{\beta,\eta}$ is common to all samples, the existence of such
a threshold depends only on the direction $\theta=(\alpha,\beta)$ in the
lifted plane. If deployment requires preserving the original zero threshold,
then the offset controlled by $\eta$ must additionally place the separated
score interval around zero; this case is handled after the constructive result
in Section~\ref{sec:theory}.

\subsection{Frozen representations}

The same algebra holds when the replaced layer follows an arbitrary frozen
representation. Let $g:\mathcal{X}\to\R^m$ be fixed and consider
\begin{equation}
F(x)=\sum_{j=1}^{m} a_j\,\sigma\bigl(g_j(x)\bigr)+b.
\label{eq:F-frozen}
\end{equation}
Replacing $y_j(x)$ by $g_j(x)$ in \eqref{eq:F-tilde-decomp} defines
$Q_g(x),H_g(x),\Phi_g(x)$. Therefore all binary feasibility and
coefficient-construction results below apply to frozen DINOv2 or
Qwen3-Embedding features followed by a single-hidden-layer ReLU MLP head.

\subsection{Multiclass heads}

For $K$ classes, let the trained ReLU logits be
\begin{equation}
F_c(x)=\sum_{j=1}^{m} a_{c,j}\,\sigma\bigl(y_j(x)\bigr)+b_c,\quad c=1,\dots,K.
\label{eq:F-mc}
\end{equation}
In the reported experiments, the target class is
$t_i=\operatorname{Top1}(F_1(x_i),\ldots,F_K(x_i))$, with a fixed deterministic
tie rule if needed. Exact preservation below is expressed with positive pairwise
logit margins; thus, when the original ReLU logits have a unique
top-1 class, these inequalities preserve that top-1 decision with positive
margin.

Define class-wise quantities
\begin{equation}
\begin{aligned}
Q_c(x)&=\sum_j a_{c,j}\,y_j(x)^2,\qquad
L_c(x)=\sum_j a_{c,j}\,y_j(x),\\
H_c(x)&=L_c(x)+b_c,
\qquad
B_c=\sum_j a_{c,j}.
\end{aligned}
\label{eq:QHc}
\end{equation}
Here $L_c$ is the bias-free linear statistic, while $H_c$ is the affine
statistic analogous to the binary notation. After the replacement,
\begin{equation}
\begin{aligned}
\widetilde F_c(x)
&=\alpha Q_c(x)+\beta L_c(x)+\eta B_c+b_c\\
&=\alpha Q_c(x)+\beta H_c(x)+\eta B_c+(1-\beta)b_c.
\end{aligned}
\label{eq:Ftilde-mc}
\end{equation}
The replacement preserves the target top-1 decision of $x_i$ exactly when
\begin{equation}
\widetilde{F}_{t_i}(x_i)>\widetilde{F}_c(x_i)\;\;\forall c\ne t_i.
\label{eq:mc-cond}
\end{equation}
For each pair $(i,c)$, where $c\ne t_i$, define target-minus-competitor
quantities
\begin{equation}
\Delta Q_{i,c}=Q_{t_i}(x_i)-Q_c(x_i),\quad
\Delta L_{i,c}=L_{t_i}(x_i)-L_c(x_i),
\end{equation}
\begin{equation}
\Delta B_{i,c}=B_{t_i}-B_c,
\qquad
\Delta b_{i,c}=b_{t_i}-b_c.
\end{equation}
Then the pairwise logit margin is the affine function
\begin{equation}
M_{i,c}(\alpha,\beta,\eta)=
\alpha\Delta Q_{i,c}+\beta\Delta L_{i,c}+\eta\Delta B_{i,c}+\Delta b_{i,c}.
\label{eq:Mic}
\end{equation}
Equivalently, since $\Delta L_{i,c}=\Delta H_{i,c}-\Delta b_{i,c}$,
\begin{equation}
M_{i,c}=\alpha\Delta Q_{i,c}+\beta\Delta H_{i,c}+(1-\beta)\Delta b_{i,c}+\eta\Delta B_{i,c}.
\end{equation}
The homogeneous multiclass optimization in Section~\ref{sec:theory} uses
the bias-free feature $\Delta L_{i,c}$ in \eqref{eq:Mic}. This avoids
double-counting the output bias: if one instead stores $\Delta H_{i,c}$
inside the lifted vector, the last coordinate must be $(\lambda-\tilde\beta)\Delta
b_{i,c}$ rather than $\lambda\Delta b_{i,c}$, where $\tilde\beta$ is the homogeneous coordinate corresponding to $\beta$ in Section~\ref{sec:theory}. We therefore use
\eqref{eq:Mic} throughout the QPs and the implementation. The multiclass exact
condition is a finite set of affine inequalities in $(\alpha,\beta,\eta)$.
Once these inequalities are infeasible, the soft-margin formulation below is
a convex surrogate for high agreement, not an exact calibration-lossless
certificate.

\begin{table}[t]
\centering
\caption{Main notation for the binary geometric formulation.}
\label{tab:notation}
\small
\begin{tabular}{@{}l p{0.72\columnwidth}@{}}
\toprule
Symbol & Meaning \\
\midrule
$\D^+, \D^-$ & positive and negative target calibration sets induced by original ReLU decisions \\
$y_j(x)$ & fixed pre-activation of hidden unit $j$ \\
$Q(x)$ & weighted quadratic statistic $\sum_j a_j\,y_j(x)^2$ \\
$H(x)$ & weighted affine statistic $\sum_j a_j\,y_j(x)+b$ \\
$L_c(x)$ & multiclass bias-free statistic $\sum_j a_{c,j}y_j(x)$ \\
$H_c(x)$ & multiclass affine statistic $L_c(x)+b_c$ (not lifted in MC-QP) \\
$B$ & output-weight sum $\sum_j a_j$ \\
$\Phi(x)$ & binary quadratic lift $(Q(x),H(x))^{\!\top}\in\R^2$ \\
$S^\pm$ & lifted point clouds $\{\Phi(x):x\in\D^\pm\}$ \\
$C^\pm$ & convex hulls $\mathrm{conv}(S^\pm)$ \\
$\mathcal{K}$ & difference hull $C^+-C^-$ \\
$m(\theta)$ & directional decision margin in direction $\theta=(\alpha,\beta)^{\!\top}$ \\
$\rho(\theta)$ & normalized margin $m(\theta)/\|\theta\|_2$ \\
\bottomrule
\end{tabular}
\end{table}

\section{Convex-Geometric Theory}
\label{sec:theory}

\subsection{Binary exact replacement}

Let
\begin{equation}
S^+=\{\Phi(x):x\in\D^+\},\quad S^-=\{\Phi(x):x\in\D^-\},
\end{equation}
and let $C^+=\mathrm{conv}(S^+),C^-=\mathrm{conv}(S^-)$. For
$\theta=(\alpha,\beta)^{\!\top}$ define the \emph{directional margin}
\begin{equation}
m(\theta)=\min_{u\in C^+}\theta^{\!\top}u-\max_{v\in C^-}\theta^{\!\top}v.
\label{eq:m-theta}
\end{equation}

\begin{theorem}[Exact binary replacement]
\label{thm:exact-binary}
Assume that both $\D^+$ and $\D^-$ are nonempty and that the deployment
threshold may be chosen after replacement. A shared quadratic replacement
$q(u;\alpha,\beta,\eta)$ is calibration-lossless for the binary target split
on $\D_{\mathrm{cal}}$ if and only if there exists
$\theta=(\alpha,\beta)\ne 0$ such that $m(\theta)>0$. The following five
conditions are equivalent:
\begin{enumerate}
\item there exists $\theta\ne 0$ with $m(\theta)>0$;
\item $C^+$ and $C^-$ admit a positive-margin hyperplane separator, i.e.,
there exist $\theta\ne0$ and $t\in\R$ such that
\begin{equation}
\min_{u\in C^+}\theta^{\!\top}u>t>\max_{v\in C^-}\theta^{\!\top}v;
\end{equation}
\item the lifted calibration sets $S^+$ and $S^-$ are positive-margin separable;
\item $C^+\cap C^-=\varnothing$;
\item $0\notin \mathcal{K}:=C^+-C^-$.\end{enumerate}
\end{theorem}

Theorem~\ref{thm:exact-binary} gives both a positive and a negative answer
on the calibration set. If the hulls are disjoint, a calibration-lossless
post-training quadratic replacement exists. If the hulls intersect, no choice
of $(\alpha,\beta,\eta)$ can separate all positive and negative target
calibration decisions under the shared quadratic replacement family, even if a
post-replacement threshold is allowed. The positive-margin separation equivalences are standard; the
contribution is the reduction of a trained ReLU-replacement problem to this
two-dimensional condition. Theorem~\ref{thm:exact-binary} assumes the
deployment threshold may be chosen after replacement; preserving the original
zero threshold is a strictly stronger requirement, handled constructively in
the fixed-zero result below (Eq.~\eqref{eq:eta-fixed}), and the reported binary
feasibility logs already enforce it.

The condition is low-dimensional but not weak. The two coordinates $Q(x)$ and
$H(x)$ are not hand-designed features; they are exactly the two
sample-dependent statistics through which a shared quadratic can affect the
binary score once the trained affine head is fixed. Theorem~\ref{thm:exact-binary}
therefore characterizes every possible post-training shared quadratic
replacement for this binary layer: overlapping lifted hulls make every affine
separator fail on some positive--negative convex combination, while disjoint
hulls yield a valid quadratic direction.

\begin{figure}[t]
\centering
\includegraphics[width=0.82\linewidth]{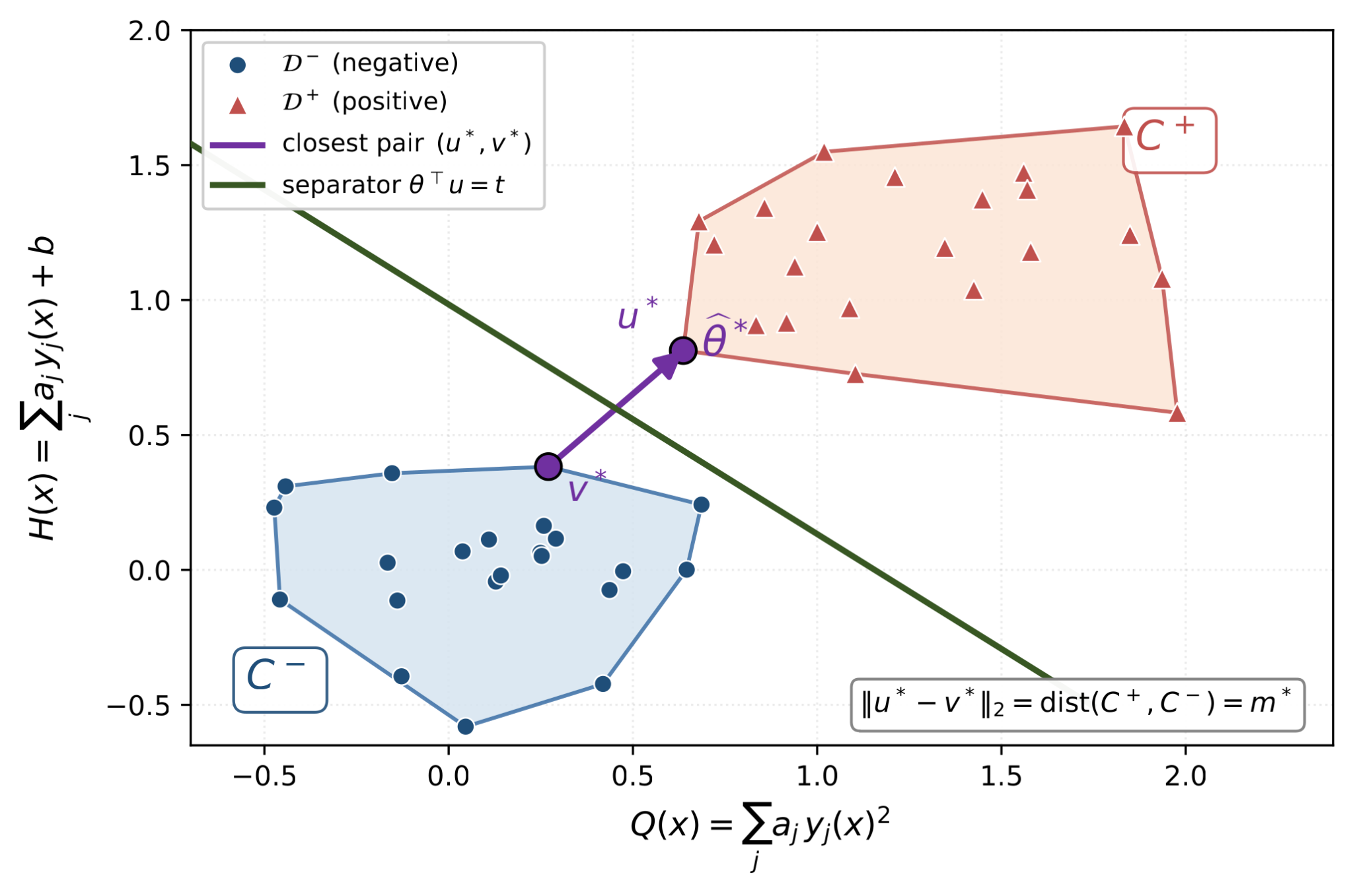}
\caption{Binary post-training quadratic replacement as positive-margin hyperplane
separation in the lifted plane. Calibration-lossless binary replacement is
possible iff the positive and negative convex hulls of $\Phi(x)=(Q(x),H(x))$
admit a positive-margin hyperplane separator. The maximum-margin direction
$\widehat{\theta}^*=(u^*-v^*)/\|u^*-v^*\|_2$ is the unit vector along the
closest-pair segment $(u^*,v^*)$, and
$m^*=\|u^*-v^*\|_2=\mathrm{dist}(C^+,C^-)$ is the optimal directional
margin.}
\label{fig:quad-geometry}
\end{figure}

Figure~\ref{fig:quad-geometry} illustrates this binary geometry: a
high-dimensional hidden layer, once the trained output weights are fixed,
reduces to two planar statistics whose hulls decide replacement feasibility.

\subsection{Constructive maximum-margin coefficients}

With $\mathcal{K}=C^+-C^-$ as in Theorem~\ref{thm:exact-binary}, $m(\theta)=\min_{z\in \mathcal{K}}\theta^{\!\top}z$. If
$0\notin \mathcal{K}$, let
\begin{equation}
z^*\in\arg\min_{z\in \mathcal{K}}\|z\|_2,
\qquad
\widehat{\theta}^*=z^*/\|z^*\|_2.
\label{eq:projection}
\end{equation}

\begin{theorem}[Maximum-margin direction]
\label{thm:max-margin}
If $C^+\cap C^-=\varnothing$, then $\widehat{\theta}^*$ maximizes the
normalized margin $\rho(\theta)=m(\theta)/\|\theta\|_2$, and the optimal
margin equals the distance between the two convex hulls:
\begin{equation}
\max_{\theta\ne 0}\frac{m(\theta)}{\|\theta\|_2}=\mathrm{dist}(C^+,C^-).
\label{eq:dist-margin}
\end{equation}
If $(u^*,v^*)\in\arg\min_{u\in C^+,v\in C^-}\|u-v\|_2$, then
$\widehat{\theta}^*=(u^*-v^*)/\|u^*-v^*\|_2$.
\end{theorem}

Because the hulls are planar, the construction is efficient. After computing
the hidden pre-activations and the lifted points, the two convex hulls can be
built in $O(n\log n)$ time; intersection testing and nearest-point computation
are linear in the number of hull vertices. When deployment must use the
original zero threshold rather than a selected post-replacement threshold, the
constant $C_{\beta,\eta}$ must place the separated score interval around zero.
If $B\ne 0$, this can be done by selecting
\begin{equation}
\eta=\frac{-(\ell^++h^-)/2-(1-\beta)b}{B},
\label{eq:eta-fixed}
\end{equation}
where $\ell^+=\min_{x\in\D^+}\theta^{\!\top}\Phi(x)$ and
$h^-=\max_{x\in\D^-}\theta^{\!\top}\Phi(x)$. If $B=0$, $\eta$ does not shift
the binary scores, so zero-threshold preservation must be checked directly.

\subsection{Quantization-tolerance certificate}

The same margin also certifies robustness to coefficient quantization, which
is essential because CKKS encodings have finite precision and deployment-time
scale.

\begin{corollary}[Quantization tolerance]
\label{cor:quantization}
Let $\widehat{\theta}^*$ be the unit-norm maximum-margin direction of
Theorem~\ref{thm:max-margin}, and let $m^*=m(\widehat{\theta}^*)>0$. Define
\begin{equation}
M=\max_{x^+\in\D^+,\,x^-\in\D^-}\|\Phi(x^+)-\Phi(x^-)\|_2.
\label{eq:M-diam}
\end{equation}
Here $M=\operatorname{diam}(\mathcal{K})$: since $\mathcal{K}=C^+-C^-$ is a
polytope, $\max_{z\in\mathcal{K}}\|z\|_2$ is attained at a vertex,\footnote{A
vertex-reduction lemma and the complete proofs of Theorems~\ref{thm:exact-binary}--\ref{thm:mc-exact}
and Corollary~\ref{cor:quantization} are given in the supplementary material.}
i.e., at a difference of extreme points of $C^\pm$, which are original
calibration samples; hence the hull-level diameter coincides with the
sample-pair maximum in~\eqref{eq:M-diam}. If a quantized coefficient vector $\widehat{\theta}$
satisfies $\|\widehat{\theta}-\widehat{\theta}^*\|_2<m^*/M$, then
$m(\widehat{\theta})>0$ and all binary calibration decisions remain preserved.
Quantitatively,
$m(\widehat{\theta})\ge m^*-M\,\|\widehat{\theta}-\widehat{\theta}^*\|_2$.
\end{corollary}

\subsection{Relaxation beyond positive-margin feasibility}

The preceding subsections characterize when a trained ReLU layer admits
calibration-lossless quadratic replacement: the lifted positive and
negative convex hulls must admit a positive-margin separator. This
condition can fail even when the original ReLU model is
accurate---because the ReLU decision margins are small, because
near-boundary or influential calibration points dominate the convex hulls,
or because multiclass pairwise constraints are mutually inconsistent.
Geometrically, such samples can bring the two hulls into contact. An immediate remedy is to discard such samples and recompute
the closest-pair construction; if only a few are removed, the problem
reduces to the positive-margin separable regime of
Theorem~\ref{thm:exact-binary}. We
avoid this hard-deletion route, because the fitted coefficients are
calibration-set dependent: aggressive removal distorts the empirical
distribution they are tuned to, and no a priori bound limits how many
samples must be discarded, especially when the failure reflects
intrinsic margin degradation rather than a few outliers. Instead, we
relax the geometric object itself, using two large-margin relaxations.
Replacing standard convex hulls with reduced convex hulls (RCH) caps the
maximum weight any single sample may carry in the convex combination,
limiting the influence of extreme points without a discrete deletion set.
As shown below, this relaxation admits an equivalent $\nu$-SVM-type
convex quadratic program and reduces to the closest-pair construction of
Theorem~\ref{thm:max-margin} at $\mu=1$, so the family indexed by $\mu$
continuously extends the exact theory rather than replacing it.

\paragraph*{Reduced convex hull (RCH)}
The first is a reduced-convex-hull (RCH) relaxation~\cite{cortes1995svm,
bennett2000svm}, a geometric large-margin relaxation that limits the influence
of any single calibration point. For a finite set $S=\{z_i\}_{i=1}^n$ and
$\mu\in[1/n,1]$, define
\begin{equation}
\mathrm{RCH}_\mu(S)=\Bigl\{\textstyle\sum_i \lambda_i z_i:\;\sum_i\lambda_i=1,\;0\le\lambda_i\le\mu\Bigr\}.
\label{eq:rch-def}
\end{equation}
Replacing $C^+,C^-$ by reduced hulls limits the effect of extreme points. For
$\mu<1$, positive-margin separation of reduced hulls is a relaxation rather than a certificate
of exact preservation on the full calibration set; after coefficients are
returned, agreement with all original ReLU decisions is measured directly.

\begin{theorem}[Positive-margin RCH separation]
\label{thm:rch}
For $\mu_+\in[1/n_+,1],\mu_-\in[1/n_-,1]$, the reduced hulls
$C^+_{\mu_+}=\mathrm{RCH}_{\mu_+}(S^+),\;C^-_{\mu_-}=\mathrm{RCH}_{\mu_-}(S^-)$
are nonempty, compact, and convex. The same five-fold equivalence of
Theorem~\ref{thm:exact-binary} holds with $C^\pm$ replaced by
$C^\pm_{\mu_\pm}$, and the maximum-margin direction is again the normalized
closest-point direction between the two reduced hulls.
\end{theorem}

\subsection{RCH duality and a \texorpdfstring{$\nu$}{nu}-SVM-type quadratic program}

Although this reduced-hull separation is geometrically clean, in practice it is often
more convenient to solve an equivalent convex QP. We make this precise. With
$S^\pm$ as in Theorem~\ref{thm:rch}, consider the primal
\begin{equation}
\begin{aligned}
\min_{\theta,t,\rho,\xi^\pm}\;
&\tfrac12\|\theta\|_2^2-\rho+c_+\textstyle\sum_i\xi_i^+ + c_-\sum_j\xi_j^-\\
\text{s.t.}\;
& \theta^{\!\top}z_i^+-t\ge\rho-\xi_i^+,
\quad \xi_i^+\ge 0,\\
& t-\theta^{\!\top}z_j^-\ge\rho-\xi_j^-,
\quad \xi_j^-\ge 0.
\end{aligned}
\tag{P-RCH}\label{eq:P-rch}
\end{equation}

\begin{theorem}[RCH--$\nu$-SVM duality]
\label{thm:rch-dual}
The Lagrangian dual of \eqref{eq:P-rch}, after the change of variables
$\lambda_i=2a_i,\kappa_j=2b_j,\mu_+=2c_+,\mu_-=2c_-$, equals the two-hull
closest-point problem
\begin{equation}
\begin{aligned}
\min_{\lambda,\kappa}\;
&\Bigl\|\textstyle\sum_i\lambda_i z_i^+-\sum_j\kappa_j z_j^-\Bigr\|_2^2\\
\text{s.t.}\;
& \sum_i\lambda_i=1,\;\sum_j\kappa_j=1,\;0\le\lambda_i\le\mu_+,\;0\le\kappa_j\le\mu_-.
\end{aligned}
\tag{D-RCH}\label{eq:D-rch}
\end{equation}
In particular, any optimal $(\theta^*,t^*,\rho^*)$ for \eqref{eq:P-rch}
satisfies $\theta^*\propto z^*=u^*-v^*$ where $u^*,v^*$ are the closest
points of $C^+_{\mu_+}$ and $C^-_{\mu_-}$.
\end{theorem}

Theorem~\ref{thm:rch-dual} unifies the geometric closest-point view of
Theorem~\ref{thm:rch} with the large-margin QP view of \eqref{eq:P-rch}; either
form may be solved in practice.

\subsection{Multiclass: pairwise margin theory}

\begin{theorem}[Multiclass exact preservation]
\label{thm:mc-exact}
Let $t_i$ denote the target class for sample $x_i$. The replacement preserves
all target top-1 decisions on $\D_{\mathrm{cal}}$ if and only if
$M_{i,c}(\alpha,\beta,\eta)>0$ for every $i$ and every $c\ne t_i$, with
$M_{i,c}$ as in \eqref{eq:Mic}. The feasible set
$\mathcal{T}_{\mathrm{mc}}=\{(\alpha,\beta,\eta):\Gamma>0\}$, where
$\Gamma=\min_{i,c\ne t_i}M_{i,c}$, is an open convex polyhedron, possibly
empty.
\end{theorem}

\paragraph*{Hard- and soft-margin formulations}
\begin{definition}[Multiclass pairwise lift]
\label{def:mc-lift}
For each sample--competitor pair $(i,c)$ with $c\ne t_i$, define the
four-dimensional pairwise lift
\begin{equation}
\bar z_{i,c}=(\Delta Q_{i,c},\Delta L_{i,c},\Delta B_{i,c},\Delta b_{i,c})^{\!\top}.
\label{eq:mc-lift-corrected}
\end{equation}
This lift stores exactly the coefficients of the affine margin
$M_{i,c}(\alpha,\beta,\eta)$ in \eqref{eq:Mic}. The last coordinate is the
fixed output-bias difference, which is why the homogeneous formulation below
uses a separate scale coordinate.
\end{definition}

Let $\bar\theta=(\tilde\alpha,\tilde\beta,\tilde\eta,\lambda)^{\!\top}\in\R^4$, where $(\tilde\alpha,\tilde\beta,\tilde\eta)$ are the homogeneous lifts of $(\alpha,\beta,\eta)$. For any $\lambda>0$,
\begin{equation}
\bar\theta^{\!\top}\bar z_{i,c}
=\lambda\,M_{i,c}\!\left(\frac{\tilde\alpha}{\lambda},\frac{\tilde\beta}{\lambda},\frac{\tilde\eta}{\lambda}\right).
\label{eq:mc-hom-equiv}
\end{equation}
Thus $\alpha=\tilde\alpha/\lambda,\;\beta=\tilde\beta/\lambda,\;\eta=\tilde\eta/\lambda$ recovers the original replacement parameters. The hard-margin feasibility QP is
\begin{equation}
\min_{\bar\theta}\;\tfrac12\|\bar\theta\|_2^2\quad
\text{s.t.}\;\bar\theta^{\!\top}\bar z_{i,c}\ge 1\;\forall i,\,c\ne t_i,\;\;\lambda\ge 1.
\tag{MC-H}\label{eq:MC-H}
\end{equation}
When \eqref{eq:MC-H} is infeasible, we solve the soft-margin variant
\begin{equation}
\begin{aligned}
\min_{\bar\theta,\xi}\;&\tfrac12\|\bar\theta\|_2^2+C\textstyle\sum_i\xi_i\\
\text{s.t.}\;&\bar\theta^{\!\top}\bar z_{i,c}\ge 1-\xi_i\;\forall i,\,c\ne t_i,\\
&\xi_i\ge 0,\;\;\lambda\ge 1.
\end{aligned}
\tag{MC-S}\label{eq:MC-S}
\end{equation}
Both \eqref{eq:MC-H} and \eqref{eq:MC-S} are convex QPs. The hard-margin
problem is an exact feasibility test for multiclass calibration-lossless
replacement under the shared quadratic family. The soft-margin problem is a
surrogate that trades margin size against violations of the original ReLU
target decisions. Implementation choices for solving these QPs at scale are
described in Section~\ref{sec:algo}; each run records the regime, the minimum
calibration margin or aggregate slack, and calibration/test agreement.

\section{Algorithms, Complexity, and FHE Cost}
\label{sec:algo}

The coefficient search is an offline post-training procedure. It uses the
trained ReLU model and a declared calibration set to compute the three shared
quadratic coefficients; the encrypted online path then evaluates only the
polynomialized network. The algorithms below summarize the binary and
multiclass cascades used in \textsc{Quad4FHE}. Complete implementation-level
pseudocode, solver settings, and logging fields are provided in the
supplementary material and the public repository.

\begin{algorithm}[H]
\caption{Binary hard/RCH/soft cascade.}
\label{alg:binary-cascade}
\small
\noindent\textbf{Input:} trained binary ReLU MLP; calibration set
$\D_{\mathrm{cal}}$; original ReLU decisions $t_i\in\{+1,-1\}$;
RCH cap grid $\mathcal M$; optional soft-margin grid.

\smallskip
\noindent\textbf{Procedure:}
\begin{enumerate}[leftmargin=1.5em,itemsep=1pt,topsep=1pt,parsep=0pt]
\item Cache hidden pre-activations $y_j(x_i)$ and compute the binary lift
      $\Phi_i=(Q_i,H_i)$ from Definition~\ref{def:binary-lift}.
\item Split lifted points into $S^+=\{\Phi_i:t_i=+1\}$ and
      $S^-=\{\Phi_i:t_i=-1\}$.
\item \emph{Hard test.} Compute the closest pair between
      $\operatorname{conv}(S^+)$ and $\operatorname{conv}(S^-)$. If their
      distance is positive, return the normalized closest-pair direction
      and the resulting quadratic coefficients with regime \textsc{hard}.
\item \emph{RCH scan.} Otherwise scan $\mu\in\mathcal M$ and test
      positive-margin separation of $\mathrm{RCH}_{\mu_+}(S^+)$ and
      $\mathrm{RCH}_{\mu_-}(S^-)$, where $\mu_\pm=\max\{\mu,1/n_\pm\}$.
      Return the first selected positive-margin RCH solution and measure
      full calibration agreement.
\item \emph{Soft fallback.} If no RCH cap separates, solve the binary
      soft-margin fallback and select coefficients by calibration
      agreement, slack, and margin diagnostics.
\end{enumerate}

\smallskip
\noindent\textbf{Output:} $(\alpha,\beta,\eta)$; regime; margin/slack
statistics; calibration agreement; mismatch indices.
\end{algorithm}

\begin{algorithm}[H]
\caption{Multiclass hard/soft-margin coefficient construction.}
\label{alg:multiclass-cascade}
\small
\noindent\textbf{Input:} trained $K$-class ReLU MLP head; calibration set
$\D_{\mathrm{cal}}$; original ReLU top-1 targets $t_i$;
soft-margin grid $\mathcal C$.

\smallskip
\noindent\textbf{Procedure:}
\begin{enumerate}[leftmargin=1.5em,itemsep=1pt,topsep=1pt,parsep=0pt]
\item Cache hidden pre-activations and compute class-wise statistics
      $Q_{i,c},L_{i,c},B_c$ and output-bias terms.
\item For each sample--competitor pair $(i,c)$ with $c\ne t_i$, form or
      stream the pairwise lift $\bar z_{i,c}$ from
      Definition~\ref{def:mc-lift}.
\item \emph{Hard QP.} Attempt the hard-margin QP~\eqref{eq:MC-H}. For
      large constraint matrices, use an active-set/cutting-plane loop:
      solve on an active subset, sweep all margins, add the worst
      violated rows, and repeat.
\item \emph{Hard recovery.} If the hard problem is feasible with positive
      margins, recover $(\alpha,\beta,\eta)$ from the homogeneous
      variables and return regime \textsc{hard}.
\item \emph{Soft fallback.} Otherwise solve the soft-margin
      problem~\eqref{eq:MC-S} over $C\in\mathcal C$, using warm starts or
      the analytically eliminated per-sample slacks. Select $C$ by
      ReLU-decision agreement, aggregate slack, worst margin, and
      coefficient norm.
\end{enumerate}

\smallskip
\noindent\textbf{Output:} $(\alpha,\beta,\eta)$; selected $C$ if
applicable; slack trace; margins; calibration/test agreement; mismatch
counts.
\end{algorithm}

\subsection{Binary complexity}

For a dense first layer, evaluating hidden pre-activations from raw inputs
costs $O(nmd)$ for $n$ calibration samples, input dimension $d$, and hidden
width $m$. Once the pre-activations are cached, computing $Q_i,H_i$ costs
$O(nm)$ and the lifted points lie in $\R^2$. We distinguish a structural
result from the executed pipeline. The $O(n\log n)$ bound is structural: it
characterizes the intrinsic difficulty of the hard positive-margin binary
case, since the planar hulls, their intersection, and the closest pair are all
computable in $O(n\log n)$. The shipped implementation, however, does not run
this planar-hull routine; it uses a shared convex-QP back-end for the hard
boundary case, the RCH scan, and the soft fallback, an engineering choice that
keeps the \textsc{hard}, \textsc{rch}, and \textsc{soft} branches numerically
consistent. For an RCH grid of size $G_\mu$, the offline cost is that of up to
$G_\mu$ two-hull closest-point QPs in two lifted dimensions. No RCH or
soft-margin optimization is part of the encrypted inference circuit.

\subsection{Multiclass and relaxed complexity}

For a $K$-class head, computing the class-wise quadratic and linear statistics
from cached hidden pre-activations costs $O(nmK)$. Each sample contributes
$K-1$ pairwise margin constraints, so a materialized multiclass constraint
matrix has $O(n(K-1))$ rows and four columns in the homogeneous formulation.
The exact multiclass problem is low-dimensional in the replacement
coefficients but may contain millions of constraints; for this reason the
solver is run offline and can stream constraints or use active sets.

In the hard-margin branch, a direct QP solve is used for moderate constraint
counts. When the number of pairwise rows exceeds the implementation threshold,
\textsc{Quad4FHE} uses the active-set/cutting-plane loop in
Algorithm~\ref{alg:multiclass-cascade}. Each outer iteration solves a small
QP on the current active set and then performs one full margin sweep costing
$O(n(K-1))$ arithmetic operations in the four-dimensional lift. In the
soft-margin branch, the per-sample slacks can be eliminated as
\begin{equation}
\xi_i(\theta)=\max\{0,\,1-\min_{c\ne t_i}M_{i,c}(\theta)\},
\end{equation}
so each objective and subgradient evaluation over a fixed $C$ again requires a
full pairwise margin sweep. The $C$ grid is selected by agreement with the
original ReLU decisions on the calibration set, then by aggregate slack,
worst margin, and coefficient norm. These costs are offline fitting costs;
the online encrypted model remains the same degree-2 arithmetic circuit for
all regimes.

\begin{table}[t]
\centering
\caption{Coefficient-construction regimes reported by \textsc{Quad4FHE}. The optimization dimension is the number of replacement coefficients, not the number of network weights.}
\label{tab:regimes}
\small
\begin{tabular}{@{}l p{0.25\linewidth} p{0.36\linewidth}@{}}
\toprule
Regime & Condition checked & Output meaning \\
\midrule
\textsc{hard} (binary)
  & $C^+\cap C^-=\varnothing$ in $\R^2$
  & exact calibration-lossless coefficients; $O(n\log n)$ after lift \\
\textsc{hard} (multiclass)
  & all pairwise affine margins feasible
  & exact positive-margin feasible coefficients \\
\textsc{rch}($\mu$)
  & reduced hulls admit a positive-margin separator
  & relaxed coefficients; full-set agreement measured afterward \\
\textsc{soft}($C$)
  & hard problem infeasible
  & slack-controlled coefficients for high empirical agreement \\
\bottomrule
\end{tabular}
\end{table}

\subsection{Encrypted arithmetic cost}

For each encrypted pre-activation $u$, \textsc{Quad4FHE} evaluates
$q(u)=\alpha u^2+\beta u+\eta$. The activation module therefore uses one
ciphertext-ciphertext multiplication to form $u^2$, followed by plaintext
multiplications, additions, and the required CKKS rescaling/relinearization
steps. In contrast, a degree-$d$ fixed-interval polynomial requires a deeper
evaluation schedule under Horner, Paterson--Stockmeyer, or related strategies,
and generally consumes more multiplication levels and coefficient-modulus
budget as $d$ increases.

The advantage of \textsc{Quad4FHE} is not that a quadratic is a universally
better pointwise approximation to ReLU. Its advantage is that the quadratic
coefficients are chosen against the trained classifier's calibration logit-ordering constraints
while preserving the circuit shape of a degree-2 activation. A shorter
multiplicative path can use a shorter coefficient-modulus chain and, within a
declared CKKS parameter-search grid, may allow a lower-latency configuration.
Section~\ref{sec:ciphertext} reports this effect relative to Remez-7; whereas against
schemes that already fit in the same depth-4 circuit, latency is similar, and
the relevant comparison is accuracy at the same encrypted depth.

\section{Experimental Methodology}
\label{sec:method}

\subsection{Models and datasets}

We evaluate two model families.

\paragraph*{Family A: single-hidden-layer MLP on raw task features.}
We test on $8$ datasets covering tabular, vision, and text modalities:
AG~News, Breast~Cancer~Wisconsin (BWC), CIFAR-10, CIFAR-100, Diabetes,
Otto~Group, Shuttle, and SST-5.

\paragraph*{Family B: frozen backbone $+$ single-hidden-layer MLP head.}
We use DINOv2 features~\cite{oquab2023dinov2} on CIFAR-100, FGVC-Aircraft,
StanfordCars, and Tiny-ImageNet, and Qwen3-Embedding-0.6B
features~\cite{qwen2025qwen3embedding} on SIB-200, MASSIVE, and Banking77.

\subsection{Splits and coefficient-fitting protocol}

For \emph{full-train} experiments, the model-training, validation, and
testing proportions are $60\%$, $20\%$, and $20\%$ when a fixed public split
is not imposed. CIFAR-10, CIFAR-100, and StanfordCars use their official
train/test partitions; only the available training side is subdivided when a
validation or calibration subset is needed. Hidden widths
$m\in\{64,128,256\}$ are evaluated; the main tables report $m=256$, which is
also used for the ciphertext experiments.

For \emph{small-pool} experiments, the model itself is trained from a $50\%$
training split, with $10\%$ for validation and $20\%$ for testing. The
remaining $20\%$ is reserved as a replacement-fitting pool, and only $1\%$,
$5\%$, $10\%$, or $20\%$ of the full dataset is used to estimate the
quadratic coefficients. Unless otherwise stated, these are single fixed
stratified splits with seed 2026 rather than repeated random splits; the
purpose is to diagnose calibration-pool size, not to estimate sampling
variance. If a tiny stratified pool misses one or more classes, the
run is reported as unavailable rather than forcing an ill-posed coefficient
fit. This protocol models deployments where the trained network is available
but the practitioner has only a small public or otherwise representative
calibration pool rather than the full training data.

All coefficient fits use original ReLU decisions as targets, and the resulting
coefficients are fixed before plaintext and CKKS evaluation. Thus
\textsc{Quad4FHE} is evaluated as a post-training agreement method rather than
as model retraining.

\subsection{Baselines}

We compare against \textbf{Square}, \textbf{least-squares polynomial fits}
of degrees $2$/$3$/$5$/$7$, \textbf{Remez polynomials} of degrees
$2$/$3$/$5$/$7$, \textbf{Precise
Approximation}~\cite{lee2023precise}, and \textbf{OLA}~\cite{lee2023ola}.
The least-squares baselines are fit to the scalar activation on a dense grid
over the empirical min--max range of hidden pre-activations in the fitting
split; the Remez baselines use the same interval with a dense-grid
minimax/Remez routine and an LP minimax fallback. No validation tuning or
percentile clipping is applied to these fixed-interval baselines, which keeps
them reproducible without claiming every interval choice is globally optimal.
OLA is evaluated as a Gaussian-weighted layerwise post-training approximation
with its configured degree/bandwidth sweep, and Precise uses its published
composed approximation with the empirical $B=\max|y|$ scaling and clipping.

The main plaintext tables report Square as the cheapest fixed polynomial,
Remez-7 as a high-degree fixed-interval baseline, and OLA/Precise as strong
post-training approximation baselines. We emphasize that OLA and Precise are
competitive in plaintext accuracy, and on several tasks match or exceed the
quadratic replacement; our claim is therefore not plaintext-accuracy
superiority. The distinguishing property is encrypted-circuit depth. OLA and
Precise attain their accuracy through higher-degree or composed polynomials
whose best CKKS use requires schedule-specific composition, modulus-chain
design, and Paterson--Stockmeyer-style evaluation, all of which raise
multiplicative depth. The thesis of this paper is sharper and narrower: a
\emph{decision-aware degree-2} replacement reaches accuracy comparable to these
stronger baselines while fitting in the lowest encrypted depth we evaluate.
Accordingly, the ciphertext tables include Square and Remez-2/3/5/7 as
depth-matched fixed-polynomial circuits; we do not reimplement OLA/Precise in
CKKS, since their value lies in approximation quality rather than in the
low-depth regime that this paper targets.

\subsection{Metrics}

For plaintext experiments we report top-1 accuracy against ground truth,
macro-F1, and \emph{agreement} of \textsc{Quad4FHE} with the original ReLU
model. We also record the returned regime, exactness on the calibration set,
margin/slack diagnostics, calibration/test agreement, coefficients, and any
coefficient quantization. For CKKS experiments we report plaintext and
ciphertext top-1 accuracy, ciphertext--plaintext mismatch counts, logit errors,
CKKS parameters $(N,\text{depth},\log Q)$, packing/batch size, and amortized
latency per sample. All evaluated ciphertext configurations are leveled CKKS
evaluations without bootstrapping.

\subsection{Encrypted-inference threat model}
\label{ssec:threat-model}

Our FHE experiments use the standard semi-honest CKKS encrypted-inference
setting. The client owns the sensitive input or feature vector and the CKKS
secret key; the server evaluates a public or server-owned classifier with
plaintext-encoded weights on ciphertexts; and the client decrypts the returned
logits. Security is therefore the usual semantic security of leveled CKKS in
this semi-honest setting. We do not claim model confidentiality against the
client, malicious-security guarantees, side-channel protection, or robustness
against malformed ciphertexts.

For raw-feature MLP tasks, the encrypted value is the task feature vector. For
DINOv2 and Qwen3 experiments, the evaluated system is encrypted MLP-head
inference on encrypted features. Privacy for raw images or text therefore
requires client-side feature extraction before encryption; otherwise FHE
protects only the head computation on the provided features.

\subsection{Relaxation hyperparameters}
\label{ssec:hyperparams}

The exact hard-margin problem is attempted first. For binary fits where the positive-margin condition fails,
we then scan the RCH cap on the fixed implementation grid
\begin{equation}
\begin{split}
\mu_{\rm base}\in\{&0.80,0.60,0.40,0.30,0.20,0.15,\\
&0.10,0.08,0.05,0.03,0.02,0.01\}.
\end{split}
\end{equation}
For the two classes we use $\mu_\pm=\max\{\mu_{\rm base},1/n_\pm\}$, so that
the reduced hulls are nonempty even on small calibration sets. We select the
largest base cap yielding a positive reduced-hull margin, with ties resolved by
validation agreement and then normalized margin; this keeps the relaxation as
close as possible to the exact hull while suppressing points that dominate the
infeasibility certificate. The Diabetes setting in
Table~\ref{tab:binary-feasibility} is selected by this rule, giving
$\mu_{\rm base}=0.01$.

For multiclass soft-margin fits we use the logarithmic grid
\begin{equation}
C\in\{10^{-3},10^{-2},10^{-1},1,10,10^2\}.
\end{equation}
The selected $C$ maximizes calibration agreement with the original ReLU
decisions; ties are broken by smaller aggregate slack $\sum_i\xi_i$, then larger
worst pairwise margin, then smaller homogeneous coefficient norm. We report
$C$, the slack-positive sample count, $\sum_i\xi_i$, and the soft-selection
trace whenever \textsc{soft}($C$) is used.

\section{Plaintext Results}
\label{sec:plaintext}

\begin{table*}[t]
\centering
\caption{Full-train single-hidden-layer MLP results at width $m=256$. Values are top-1 accuracies in \%; ``Agr.'' is the agreement of \textsc{Quad4FHE} with the original ReLU model on the test split, and ``MF1'' is the macro-F1 of \textsc{Quad4FHE}. Best non-ReLU value per row is in \textbf{bold}; second-best is \underline{underlined}.}
\label{tab:mlp-fulltrain}
\small
\setlength{\tabcolsep}{8pt}
\begin{tabular}{@{}lcccccccc@{}}
\toprule
Dataset & ReLU & Sq. & Rmz-7 & OLA & Prec. & Quad. & Agr. & MF1 \\
\midrule
AG News & 89.83 & 73.30 & 89.73 & 89.80 & \underline{89.82} & \textbf{89.95} & 98.64 & 89.94 \\
CIFAR-10 & 52.14 & 35.84 & 49.68 & \underline{52.25} & 52.13 & \textbf{52.64} & 87.41 & 52.31 \\
CIFAR-100 & 20.53 & 12.03 & 18.95 & 20.52 & \underline{20.53} & \textbf{20.71} & 82.29 & 19.53 \\
Otto & 79.71 & 38.53 & 71.87 & \underline{77.74} & \textbf{79.72} & 76.05 & 87.05 & 69.36 \\
Shuttle & 99.69 & 82.56 & 84.21 & 88.55 & \textbf{99.33} & \underline{97.04} & 97.30 & 49.32 \\
SST-5 & 36.43 & 29.19 & 36.33 & \textbf{36.43} & \underline{36.43} & 36.43 & 92.31 & 31.41 \\
\bottomrule
\end{tabular}
\end{table*}

\begin{table}[t]
\centering
\caption{Binary feasibility / coefficient summary at $m=256$ from the
full-train logs. ``Margin'' is the empirical decision margin on the
training split. \textsc{rch}($\mu$) means the reduced-convex-hull regime
with cap parameter $\mu$. Both rows use the fixed-zero threshold and
record calibration/test agreement in Table~\ref{tab:decision-diagnostics}.}
\label{tab:binary-feasibility}
\small
\setlength{\tabcolsep}{4pt}
\begin{tabular}{@{}lcrrrr@{}}
\toprule
Dataset & Regime & $\alpha$ & $\beta$ & $\eta$ & Margin \\
\midrule
BWC      & \textsc{hard}                  & 0.397 & 0.918 & $+0.771$ & 0.548 \\
Diabetes & \textsc{rch}($\mu{=}0.010$)    & 0.110 & 0.994 & $-5.202$ & 0.452 \\
\bottomrule
\end{tabular}
\end{table}

\begin{table*}[t]
\centering
\caption{Decision-agreement and regime diagnostics for the full-train width-$256$ runs. $n_c$ is the calibration size and $|\mathcal P|$ the number of multiclass pairwise constraints (binary rows have none). ``Cal./Test Agr.'' are agreement with the original ReLU model, not ground-truth accuracy. ``Norm. marg.'' is the normalized empirical decision margin; for \textsc{soft} rows negative values are accompanied by the per-sample slack statistics $\#\xi_i{>}0$ and $\sum_i\xi_i$.}
\label{tab:decision-diagnostics}
\scriptsize
\setlength{\tabcolsep}{2.4pt}
\begin{tabular}{@{}l l r r c r r r r r r r@{}}
\toprule
Task & Regime & $n_c$ & $|\mathcal P|$ & Exact & Cal. Agr. & Test Agr. & Cal. mis & Test mis & Norm. marg. & $\#\xi_i{>}0$ & $\sum_i\xi_i$ \\
 & & & & & (\%) & (\%) & & & & & \\
\midrule
BWC & \textsc{hard} & 455 & -- & Y & 100.00 & 100.00 & 0 & 0 & 0.55 & -- & -- \\
Diabetes & \textsc{rch}(0.010) & 7630 & -- & N & 99.38 & 99.48 & 47 & 10 & 0.45 & -- & -- \\
\midrule
AG News & \textsc{soft}(100.0) & 96000 & 288k & N & 98.40 & 98.64 & 1535 & 327 & -3.93 & 3800 & 3800.2 \\
CIFAR-10 & \textsc{soft}(1.0) & 50000 & 450k & N & 88.74 & 87.41 & 5630 & 1259 & -5.85 & 14337 & 14328.1 \\
CIFAR-100 & \textsc{soft}(0.1) & 50000 & 4.95M & N & 83.01 & 82.29 & 8494 & 1771 & -10.89 & 20762 & 20636.8 \\
Otto & \textsc{soft}(0.001) & 49502 & 396k & N & 86.13 & 87.05 & 6864 & 1603 & -107.71 & 17702 & 17450.7 \\
Shuttle & \textsc{soft}(0.001) & 46400 & 278k & N & 97.26 & 97.30 & 1272 & 313 & -549.98 & 6081 & 5822.9 \\
SST-5 & \textsc{soft}(1.0) & 9645 & 39k & N & 92.63 & 92.31 & 711 & 170 & -0.57 & 1884 & 1764.0 \\
\midrule
DINOv2/CIFAR-100 & \textsc{soft}(1.0) & 48000 & 4.75M & N & 99.14 & 98.76 & 411 & 149 & -1.64 & 969 & 967.0 \\
DINOv2/FGVC & \textsc{soft}(10.0) & 8000 & 792k & N & 97.21 & 94.55 & 223 & 109 & -0.51 & 531 & 526.1 \\
DINOv2/Cars & \textsc{soft}(10.0) & 12948 & 2.52M & N & 98.49 & 95.74 & 196 & 138 & -1.93 & 469 & 470.6 \\
DINOv2/Tiny & \textsc{soft}(100.0) & 88000 & 17.51M & N & 98.60 & 98.47 & 1234 & 337 & -0.98 & 2905 & 2906.0 \\
\midrule
Qwen3/SIB-200 & \textsc{soft}(10.0) & 4800 & 29k & N & 99.42 & 99.26 & 28 & 9 & -0.80 & 73 & 72.2 \\
Qwen3/MASSIVE & \textsc{soft}(10.0) & 81282 & 4.80M & N & 98.58 & 98.69 & 1152 & 233 & -0.89 & 2822 & 2819.1 \\
Qwen3/Banking77 & \textsc{soft}(0.1) & 9993 & 759k & N & 99.19 & 98.47 & 81 & 47 & -1.64 & 247 & 218.3 \\
\bottomrule
\end{tabular}
\end{table*}

\subsection{Interpreting the regime diagnostics}

Table~\ref{tab:decision-diagnostics} is central to interpreting the experiments.
Only BWC is in the theorem-backed \textsc{hard} exact regime in the full-train
main runs; Diabetes requires the RCH relaxation; and every full-train multiclass
benchmark is in \textsc{soft}($C$). The multiclass accuracy numbers should
therefore not be read as evidence that exact calibration-lossless quadratic
replacement is generally feasible; they show instead that the soft-margin
surrogate can often recover high empirical agreement despite negative
worst-case pairwise margins. The calibration/test columns quantify the
extrapolation gap: on many tasks the two agreements are close, supporting the
empirical use of the calibration pool.

\subsection{Single-hidden-layer MLPs}

Table~\ref{tab:mlp-fulltrain} reports the main MLP results at width
$256$. \textsc{Quad4FHE} is the best (or tied for best) non-ReLU
replacement on $4$ of $6$ multiclass tasks (AG News, CIFAR-10, CIFAR-100,
SST-5), and matches or exceeds the original ReLU model on all four
(per-task gaps Quad$-$ReLU range from $0.00$ to $+0.50$\,pp).
By contrast, on Otto, exact multiclass preservation is infeasible and the soft-margin relaxation loses $3.66$\,pp; on Shuttle the corresponding gap is $2.65$\,pp. Even in these two harder cases, \textsc{Quad4FHE}
substantially exceeds Square (by $37.52$\,pp on Otto and $14.48$\,pp on
Shuttle) and Remez-7 (by $4.18$\,pp on Otto, $12.83$\,pp on Shuttle). OLA
and Precise are strong post-training approximation baselines; their
near-ReLU accuracy on several tasks highlights that the comparison is
not merely between low and high polynomial degree, but between
\emph{interval-level activation approximation} and \emph{task-level
decision agreement}. BWC and Diabetes are reported separately in
Table~\ref{tab:binary-feasibility} for coefficient interpretability; their
diagnostics also appear in Table~\ref{tab:decision-diagnostics}.

Two rows require careful reading. On SST-5, Quad and ReLU have nearly identical
top-1 accuracy even though test agreement is only $92.31\%$; this is possible
because changed predictions can include both newly correct and newly incorrect
samples. Agreement is therefore the appropriate decision-agreement metric,
not accuracy alone. On Shuttle, top-1 accuracy remains high because the dataset
is highly imbalanced, but the Quad macro-F1 is low; top-1 therefore overstates
minority-class behavior. We report macro-F1 in the table and do not claim that
Quad is minority-class preserving on Shuttle.

The binary feasibility logs in Table~\ref{tab:binary-feasibility}
illustrate the role of the exact theorem (Theorem~\ref{thm:exact-binary}).
BWC is hard-margin feasible at all widths; therefore the replacement
parameters come from the exact geometric condition. Diabetes is not
hard-margin feasible under the full-train logs but obtains a positive-margin RCH solution after the
RCH relaxation (Theorem~\ref{thm:rch}) at $\mu=0.01$, which is consistent
with the presence of noisy or influential extreme points in the diabetes
patient features.

\begin{table}[t]
\centering
\caption{Small-pool detailed results at width $m=256$. Entries are \textsc{Quad4FHE} top-1 accuracy in \%; the rightmost column is the original ReLU accuracy in the same split. All rows use one fixed stratified split with seed 2026. ``--'' marks fits where the requested pool is unusable because class imbalance leaves at least one class absent (e.g., BWC and DINOv2/StanfordCars at $1\%$).}
\label{tab:smallpool-detail}
\small
\setlength{\tabcolsep}{4pt}
\begin{tabular}{@{}lccccc@{}}
\toprule
Task & $1\%$ & $5\%$ & $10\%$ & $20\%$ & ReLU \\
\midrule
\multicolumn{6}{l}{\emph{Single-hidden-layer MLP}} \\
AG News & 89.95 & 90.09 & 90.15 & 90.12 & 89.99 \\
BWC & -- & 91.23 & 98.25 & 96.49 & 99.12 \\
CIFAR-10 & 49.38 & 49.45 & 49.58 & 49.84 & 50.32 \\
CIFAR-100 & 18.66 & 18.67 & 18.65 & 18.65 & 18.86 \\
Diabetes & 95.07 & 98.64 & 98.79 & 98.43 & 98.85 \\
Otto & 76.46 & 76.39 & 76.28 & 76.51 & 79.72 \\
Shuttle & 98.72 & 98.64 & 96.22 & 96.91 & 99.74 \\
SST-5 & 34.30 & 34.48 & 34.66 & 34.43 & 34.71 \\
\midrule
\multicolumn{6}{l}{\emph{DINOv2 backbone $+$ MLP head}} \\
DINOv2/CIFAR-100 & 89.52 & 89.42 & 89.38 & 89.38 & 89.41 \\
DINOv2/FGVC & 76.60 & 76.40 & 76.60 & 76.65 & 77.75 \\
DINOv2/Cars & -- & 85.51 & 85.63 & 85.60 & 85.70 \\
DINOv2/Tiny & 86.23 & 86.20 & 86.20 & 86.21 & 86.20 \\
\midrule
\multicolumn{6}{l}{\emph{Qwen3 backbone $+$ MLP head}} \\
Qwen3/SIB-200 & 93.86 & 94.44 & 94.44 & 94.44 & 94.77 \\
Qwen3/MASSIVE & 89.40 & 89.45 & 89.49 & 89.42 & 89.90 \\
Qwen3/Banking77 & 90.44 & 90.86 & 90.90 & 90.74 & 90.63 \\
\bottomrule
\end{tabular}
\end{table}

\subsection{Small-pool replacement}

Table~\ref{tab:smallpool-detail} reports the small-pool ablation. Across this
fixed split, the quadratic replacement remains usable even when only a small
fraction of the dataset is available for coefficient fitting; frozen-head
settings are typically more stable than raw-feature MLPs. These results are
calibration-pool sensitivity diagnostics under one stratified split rather than
repeated-seed statistical claims.

\begin{table*}[t]
\centering
\caption{Frozen backbone $+$ single-hidden-layer MLP head at width $m=256$. Values are top-1 accuracies in \%; ``Agr.'' is \textsc{Quad4FHE} agreement with the original ReLU head.}
\label{tab:frozen-fulltrain}
\small
\setlength{\tabcolsep}{8pt}
\begin{tabular}{@{}lccccccc@{}}
\toprule
Task & ReLU & Sq. & Rmz-7 & OLA & Prec. & Quad. & Agr. \\
\midrule
DINOv2/CIFAR-100 & 89.53 & 78.03 & \textbf{89.59} & 89.56 & 89.53 & \underline{89.57} & 98.76 \\
DINOv2/FGVC & 79.40 & 25.50 & 78.75 & \textbf{79.40} & \underline{79.40} & 78.25 & 94.55 \\
DINOv2/Cars & 87.30 & 36.92 & \underline{87.33} & 87.33 & 87.30 & \textbf{87.49} & 95.74 \\
DINOv2/Tiny & 86.48 & 70.75 & \textbf{86.54} & 86.46 & 86.47 & \underline{86.52} & 98.47 \\
Qwen3/SIB-200 & 89.30 & 52.53 & \textbf{89.79} & 89.30 & 89.30 & \underline{89.71} & 99.26 \\
Qwen3/MASSIVE & 85.56 & 52.40 & 85.49 & \textbf{85.56} & \underline{85.56} & 85.45 & 98.69 \\
Qwen3/Banking77 & 91.91 & 71.07 & 91.68 & \underline{91.84} & \textbf{91.91} & 91.55 & 98.47 \\
\midrule
\multicolumn{8}{l}{\emph{Average $|\text{Quad}-\text{ReLU}|$ gap across the 7 tasks: 0.33\,pp.}} \\
\bottomrule
\end{tabular}
\end{table*}

\subsection{Frozen representations with MLP heads}

Table~\ref{tab:frozen-fulltrain} evaluates the head-only
frozen-representation extension. \textsc{Quad4FHE} is within $0.41$\,pp of the
original ReLU head on $6$ of $7$ tasks and within $1.15$\,pp on all $7$, an
average absolute gap of $0.33$\,pp. OLA and Precise are also close to the ReLU
heads here, while Square remains much less stable. On DINOv2/StanfordCars
\textsc{Quad4FHE} is the \emph{best} non-ReLU replacement, exceeding the ReLU
head by $+0.19$\,pp. These results confirm that the coefficient construction is
not tied to raw-input MLPs; it applies whenever a fixed feature extractor is
followed by a single-hidden-layer ReLU head and an affine classifier. The
privacy claim for DINOv2/Qwen3 remains head-only encrypted inference on
encrypted features, not full encrypted image/text inference.

\section{Ciphertext Results and Analysis}
\label{sec:ciphertext}

We report CKKS inference on three representative tasks: Otto (tabular MLP), DINOv2/FGVC-Aircraft, and Qwen3/MASSIVE.

\subsection{CKKS implementation and parameter search}
\label{ssec:ckks-impl}

The ciphertext implementation uses Microsoft SEAL~4.1 with CKKS on an Intel Xeon Gold~6530 and $32$ threads for the main latency runs. The server evaluates plaintext weights on ciphertexts, the client holds the secret key, and no bootstrapping is used. For the main $\log_2\Delta=40$ runs, the coefficient-modulus chains are
\begin{equation}
\begin{array}{ll}
\text{depth }4:\;&\{60,40,40,40,40,60\}\quad(\log Q=280),\\
\text{depth }5:\;&\{60,40,40,40,40,40,60\}\quad(\log Q=320),\\
\text{depth }6:\;&\{60,40,40,40,40,40,40,60\}\quad(\log Q=360),
\end{array}
\end{equation}
and the precision sweep $\log_2\Delta\in\{25,30,35,40,45\}$ adjusts the
middle primes consistently. All configurations instantiate the SEAL context
with \texttt{sec\_level\_type::tc128} and pass the realized-level check, with
the total modulus kept below SEAL's 128-bit maximum for each ring degree;
$N=2^{15}$ is used in the larger-resource configurations.

Packing interleaves $B=\texttt{slot\_count}/m$ samples per hidden coordinate
($B=32$ at $N=2^{14}$, $B=64$ at $N=2^{15}$, $m=256$); features larger
than $m$ are chunked, each chunk evaluated by a baby-step/giant-step
diagonal matvec, and the chunk results accumulated. Latencies are amortized
per-sample (sequential-equivalent); Table~\ref{tab:ckks-opcounts} reports
per-batch operation counts.

\begin{table}[t]
\centering
\caption{Theoretical CKKS operation counts per encrypted batch for the selected
full-sweep configurations. ``ct-ct'' and ``ct-pt'' are ciphertext-ciphertext and
ciphertext-plaintext multiplications. Rotations include both BSGS input
rotations and output-tree reductions.}
\label{tab:ckks-opcounts}
\scriptsize
\setlength{\tabcolsep}{3pt}
\begin{tabular}{@{}llrrrrr@{}}
\toprule
Task & Scheme & Enc & ct-ct & ct-pt & Rot. & Rescale \\
\midrule
Otto & Quad & 1 & 2 & 139 & 93 & 14 \\
Otto & Remez-7 & 1 & 5 & 141 & 93 & 19 \\
FGVC & Quad & 3 & 2 & 870 & 890 & 105 \\
FGVC & Remez-7 & 3 & 5 & 872 & 890 & 110 \\
MASSIVE & Quad & 4 & 2 & 1086 & 600 & 65 \\
MASSIVE & Remez-7 & 4 & 5 & 1088 & 600 & 70 \\
\bottomrule
\end{tabular}
\end{table}

A configuration is called \emph{minimum feasible} only relative to the
explicit search grid
\begin{equation}
\begin{split}
(N,D_{\mathrm{depth}},\log Q)\in\{&(2^{14},4,280),(2^{14},5,320),\\
&(2^{15},5,320),(2^{15},6,360)\}.
\end{split}
\end{equation}
Here $D_{\mathrm{depth}}$ denotes the available multiplicative-depth budget
in the declared search grid. A scheme is feasible on a row if its compile-time depth does not exceed the
available rescale budget, inference completes without bootstrapping, and the
ciphertext top-1 predictions match the plaintext polynomialized top-1
predictions on every evaluated test sample. Thus ``HE--Plain=0'' below means
zero top-1 mismatches, not zero numerical logit error; max/mean logit errors
are logged separately.

\subsection{Minimum feasible CKKS configuration}

\begin{table*}[t]
\centering
\caption{Minimum feasible CKKS configuration from the ciphertext logs
(\textsc{Quad4FHE} rows are shaded). \emph{PH-mis} is the integer number
of top-1 mismatches between plaintext polynomial evaluation and CKKS evaluation on the evaluated
test set; it is zero for every selected row, so a single \textit{Acc.} column
reports both plaintext and ciphertext accuracy. \emph{Lat-Act} is the
per-sample cost of the polynomial-evaluation block (Powers $+$ Activation);
\emph{Lat-Tot} is the end-to-end per-sample latency. \emph{Spd-Act} and
\emph{Spd-Tot} are the corresponding speedups versus Remez-7. Latencies are
amortized per-sample milliseconds (sequential-equivalent, 32-thread full sweep).}
\label{tab:ckks-minconfig}
\scriptsize
\renewcommand{\arraystretch}{1.0}
\setlength{\tabcolsep}{3pt}
\begin{tabular}{@{}llccccc rrcc@{}}
\toprule
Task & Activation & Deg. & Acc.\,(\%) & PH-mis & Depth & $\log Q$ &
\multicolumn{1}{c}{Lat-Act} & \multicolumn{1}{c}{Lat-Tot} &
Spd-Act & Spd-Tot \\
 &  &  &  &  &  &  & (ms) & (ms) & vs Rmz-7 & vs Rmz-7 \\
\midrule
\multicolumn{11}{l}{\textbf{Otto / MLP}} \\
\rowcolor{gray!12} & \textsc{Quad4FHE} & 2 & 76.05 & 0 & 4 & 280 & 0.219 & 7.43 & $3.77\times$ & $1.25\times$ \\
& Square  & 2 & 38.53 & 0 & 4 & 280 & 0.180 & 7.30 & $4.59\times$ & $1.27\times$ \\
& Remez-2 & 2 & 60.59 & 0 & 4 & 280 & 0.210 & 7.30 & $3.93\times$ & $1.27\times$ \\
& Remez-3 & 3 & 61.10 & 0 & 4 & 280 & 0.309 & 7.42 & $2.67\times$ & $1.25\times$ \\
& Remez-5 & 5 & 68.63 & 0 & 5 & 320 & 0.670 & 9.10 & $1.23\times$ & $1.02\times$ \\
& Remez-7 & 7 & 71.87 & 0 & 5 & 320 & 0.826 & 9.27 & $1.00\times$ & $1.00\times$ \\
\midrule
\multicolumn{11}{l}{\textbf{DINOv2 / FGVC-Aircraft}} \\
\rowcolor{gray!12} & \textsc{Quad4FHE} & 2 & 78.25 & 0 & 4 & 280 & 0.222 & 46.81 & $4.11\times$ & $1.68\times$ \\
& Square  & 2 & 25.50 & 0 & 4 & 280 & 0.167 & 46.39 & $5.46\times$ & $1.70\times$ \\
& Remez-2 & 2 & 73.80 & 0 & 4 & 280 & 0.202 & 46.49 & $4.51\times$ & $1.70\times$ \\
& Remez-3 & 3 & 75.70 & 0 & 4 & 280 & 0.314 & 46.53 & $2.90\times$ & $1.69\times$ \\
& Remez-5 & 5 & 78.10 & 0 & 5 & 320 & 0.771 & 79.24 & $1.18\times$ & $0.99\times$ \\
& Remez-7 & 7 & 78.75 & 0 & 5 & 320 & 0.912 & 78.83 & $1.00\times$ & $1.00\times$ \\
\midrule
\multicolumn{11}{l}{\textbf{Qwen3 / MASSIVE}} \\
\rowcolor{gray!12} & \textsc{Quad4FHE} & 2 & 85.45 & 0 & 4 & 280 & 0.231 & 50.05 & $3.74\times$ & $1.18\times$ \\
& Square  & 2 & 52.40 & 0 & 4 & 280 & 0.186 & 49.68 & $4.64\times$ & $1.19\times$ \\
& Remez-2 & 2 & 83.51 & 0 & 4 & 280 & 0.217 & 49.35 & $3.98\times$ & $1.19\times$ \\
& Remez-3 & 3 & 85.14 & 0 & 4 & 280 & 0.318 & 49.55 & $2.71\times$ & $1.19\times$ \\
& Remez-5 & 5 & 85.40 & 0 & 5 & 320 & 0.711 & 58.75 & $1.21\times$ & $1.00\times$ \\
& Remez-7 & 7 & 85.49 & 0 & 5 & 320 & 0.863 & 58.96 & $1.00\times$ & $1.00\times$ \\
\bottomrule
\end{tabular}
\end{table*}

Table~\ref{tab:ckks-minconfig} reports each scheme's minimum feasible
CKKS configuration within our declared search grid on the three tasks. Across
all three tasks, \textsc{Quad4FHE}, Square, and Remez-2/Remez-3 remain feasible
at $N=2^{14}$, depth $4$, and $\log Q=280$, whereas Remez-5 and Remez-7 require
depth $5$ at $\log Q=320$. The phrase ``minimum feasible'' is therefore
relative to this grid, scale policy, packing strategy, and accuracy criterion;
it is not a proof of global optimality over all CKKS schedules.

The latency comparison should be read in two layers. First, at the same depth-4
configuration, \textsc{Quad4FHE} has latency comparable to Remez-2/3 (e.g.,
Otto: $7.43$\,ms vs $7.30$--$7.42$\,ms), so the quadratic is not intrinsically
much faster than every other low-degree circuit; its advantage is that the
decision-aware quadratic retains higher task accuracy than the low-degree
interval fits while still fitting in the lowest evaluated depth. Second, relative to
Remez-7, which often offers the strongest fixed-interval accuracy among the
CKKS baselines but requires depth $5$, the activation-block speedup is
$3.7$--$4.1\times$ across the three tasks and the end-to-end speedup is
$1.18$--$1.68\times$ (full numbers in Table~\ref{tab:ckks-minconfig});
\textsc{Quad4FHE} leads Remez-7 by $+4.18$\,pp on Otto, is within
$0.50$\,pp on DINOv2/FGVC, and matches Remez-7 within $0.04$\,pp on
Qwen3/MASSIVE.

\begin{figure}[t]
\centering
\includegraphics[width=\linewidth]{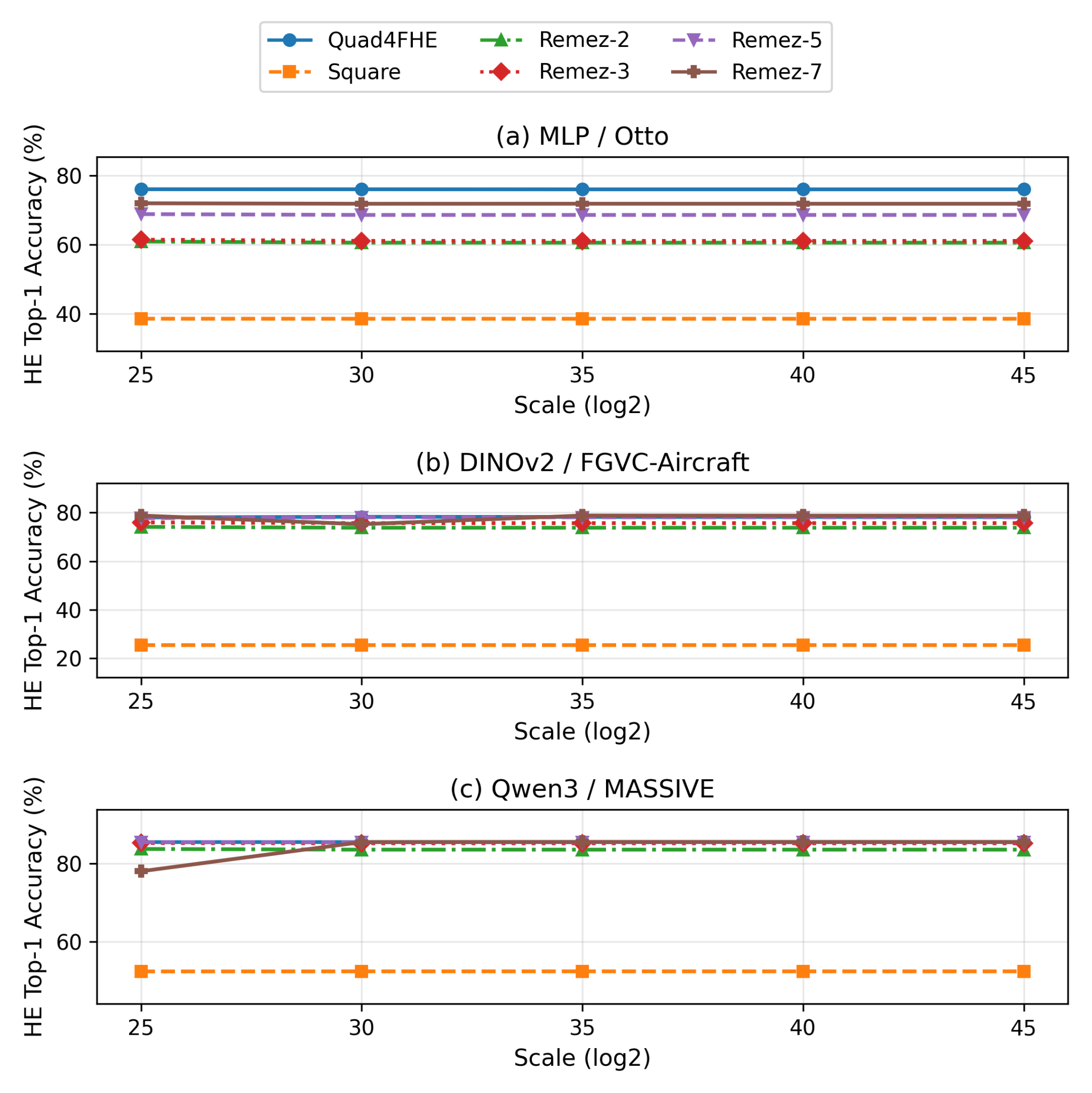}
\caption{HE top-1 accuracy versus the CKKS scaling factor $\log_2\Delta$ on the three encrypted tasks.}
\label{fig:ckks-scale-3task}
\end{figure}

\subsection{Precision sweep}

Figure~\ref{fig:ckks-scale-3task} shows the HE accuracy of all six
schemes as the CKKS scaling factor $\log_2\Delta$ ranges from $25$ to
$45$. Except for the Remez-7 dip on MASSIVE at $\log_2\Delta=25$,
accuracy is essentially flat across the precision range for the evaluated
schemes and tasks, confirming that the polynomialized circuits are
numerically stable under the tested CKKS scales. The overall ordering is
stable at moderate and high scales: \textsc{Quad4FHE} leads on Otto, while
Remez-7 has a small plaintext-level edge on FGVC and MASSIVE when
sufficient scale is used, at the cost of the depth-5 circuit. The visible
Remez-7 dip at the lowest MASSIVE scale is consistent with high-degree
polynomials being more sensitive to large-magnitude pre-activations.

\section{Discussion and Limitations}
\label{sec:discussion}

The exact theorems certify calibration-set preservation only. Table
\ref{tab:decision-diagnostics} shows that exact feasibility is rare in the
full-train benchmarks: BWC falls into the \textsc{hard} regime, Diabetes uses
RCH, and multiclass tasks use \textsc{soft}($C$). The multiclass results are
therefore evidence for a practical decision-aware surrogate rather than generic
exact preservation. Repeated-seed statistics, per-class F1, and
low-margin/outlier sensitivity remain useful follow-up diagnostics.

The encrypted experiments compare depth-matched fixed-polynomial circuits under
one packing and modulus-search framework, isolating the effect of
multiplicative depth. Optimized CKKS schedules for OLA/Precise, polynomial-only
training, bootstrapped deeper networks, and lookup/comparison-based protocols
are outside this controlled comparison. The DINOv2 and Qwen3 experiments are
encrypted MLP-head inference on encrypted features; they do not protect raw
images or text unless feature extraction is performed on the client.

\section{Conclusion}
\label{sec:conclusion}

We presented a decision-aware framework for HE-friendly quadratic ReLU
replacement, implemented in the open-source \textsc{Quad4FHE} library. The
framework shifts activation replacement from module-local approximation
fidelity to calibration-set logit-ordering agreement. For binary
single-hidden-layer MLPs, exact calibration-lossless preservation is equivalent
to positive-margin hyperplane separation of two lifted convex hulls, yielding
constructive coefficients in $O(n\log n)$ time and a quantization certificate;
for harder binary and multiclass settings, RCH and soft-margin relaxations
provide practical surrogates. CKKS experiments confirm ciphertext accuracies
matching plaintext polynomial accuracies and $1.18$--$1.68\times$ end-to-end
speedups over Remez-7.

\bibliographystyle{IEEEtran}

\begin{thebibliography}{99}
\scriptsize
\setlength{\itemsep}{0pt}
\setlength{\parskip}{0pt}

\bibitem{rivest1978privacy}
R.~L. Rivest, L.~Adleman, and M.~L. Dertouzos,
``On data banks and privacy homomorphisms,''
in \emph{Foundations of Secure Computation}, 1978, pp.~169--179.

\bibitem{gentry2009fhe}
C.~Gentry,
``A fully homomorphic encryption scheme,''
Ph.D. dissertation, Stanford University, 2009.

\bibitem{cheon2017ckks}
J.~H. Cheon, A.~Kim, M.~Kim, and Y.~Song,
``Homomorphic encryption for arithmetic of approximate numbers,''
in \emph{Proc. ASIACRYPT}, 2017, pp.~409--437.

\bibitem{bost2015ml}
R.~Bost, R.~A. Popa, S.~Tu, and S.~Goldwasser,
``Machine learning classification over encrypted data,''
in \emph{Proc. NDSS}, 2015.

\bibitem{mohassel2017secureml}
P.~Mohassel and Y.~Zhang,
``SecureML: A system for scalable privacy-preserving machine learning,''
in \emph{Proc. IEEE S\&P}, 2017, pp.~19--38.

\bibitem{liu2017minionn}
J.~Liu, M.~Juuti, Y.~Lu, and N.~Asokan,
``Oblivious neural network predictions via MiniONN transformations,''
in \emph{Proc. ACM CCS}, 2017, pp.~619--631.

\bibitem{giladbachrach2016cryptonets}
R.~Gilad-Bachrach, N.~Dowlin, K.~Laine, K.~Lauter, M.~Naehrig, and J.~Wernsing,
``CryptoNets: Applying neural networks to encrypted data with high throughput and accuracy,''
in \emph{Proc. ICML}, PMLR, vol.~48, pp.~201--210, 2016.

\bibitem{brutzkus2019lola}
A.~Brutzkus, R.~Gilad-Bachrach, and O.~Elisha,
``Low latency privacy preserving inference,''
in \emph{Proc. ICML}, 2019, pp.~812--821.

\bibitem{dathathri2019chet}
R.~Dathathri \emph{et al.},
``CHET: an optimizing compiler for fully-homomorphic neural-network inferencing,''
in \emph{Proc. ACM PLDI}, 2019, pp.~142--156.

\bibitem{dathathri2020eva}
R.~Dathathri \emph{et al.},
``EVA: An encrypted vector arithmetic language and compiler for efficient homomorphic computation,''
in \emph{Proc. ACM PLDI}, 2020, pp.~546--561.

\bibitem{boemer2019ngraph}
F.~Boemer, A.~Costache, R.~Cammarota, and C.~Wierzynski,
``nGraph-HE2: A high-throughput framework for neural network inference on encrypted data,''
in \emph{Proc. WAHC}, 2019, pp.~45--56.

\bibitem{juvekar2018gazelle}
C.~Juvekar, V.~Vaikuntanathan, and A.~Chandrakasan,
``GAZELLE: A low latency framework for secure neural network inference,''
in \emph{Proc. USENIX Security}, 2018, pp.~1651--1669.

\bibitem{mishra2020delphi}
P.~Mishra \emph{et al.},
``DELPHI: A cryptographic inference service for neural networks,''
in \emph{Proc. USENIX Security}, 2020, pp.~2505--2522.

\bibitem{rathee2020cryptflow2}
D.~Rathee \emph{et al.},
``CrypTFlow2: Practical 2-party secure inference,''
in \emph{Proc. ACM CCS}, 2020, pp.~325--342.

\bibitem{riazi2019xonn}
M.~S. Riazi \emph{et al.},
``XONN: XNOR-based oblivious deep neural network inference,''
in \emph{Proc. USENIX Security}, 2019, pp.~1501--1518.

\bibitem{huang2022cheetah}
Z.~Huang \emph{et al.},
``Cheetah: Lean and fast secure two-party deep neural network inference,''
in \emph{Proc. USENIX Security}, 2022, pp.~809--826.

\bibitem{trefethen2013}
L.~N. Trefethen,
\emph{Approximation Theory and Approximation Practice}.\hskip 1em plus 0.5em
minus 0.4em\relax SIAM, 2013.

\bibitem{boyd2004convex}
S.~Boyd and L.~Vandenberghe,
\emph{Convex Optimization}. Cambridge Univ. Press, 2004.

\bibitem{hesamifard2017cryptodl}
E.~Hesamifard, H.~Takabi, and M.~Ghasemi,
``CryptoDL: Deep neural networks over encrypted data,''
arXiv:1711.05189, 2017.

\bibitem{lee2023precise}
J.~Lee, E.~Lee, J.-W.~Lee, Y.~Kim, Y.-S.~Kim, and J.-S.~No,
``Precise approximation of convolutional neural networks for homomorphically
encrypted data,''
\emph{IEEE Access}, vol.~11, pp.~62062--62076, 2023,
doi: 10.1109/ACCESS.2023.3287564.

\bibitem{lee2023ola}
J.~Lee, E.~Lee, Y.-S.~Kim, Y.~Lee, J.-W.~Lee, Y.~Kim, and J.-S.~No,
``Optimized layerwise approximation for efficient private inference on fully
homomorphic encryption,''
arXiv:2310.10349v4, 2025.

\bibitem{lou2021safenet}
Q.~Lou, Y.~Shen, H.~Jin, and L.~Jiang,
``SAFENet: A secure, accurate, and fast neural network inference,''
in \emph{Proc. ICLR}, 2021.

\bibitem{ao2024autofhe}
W.~Ao and V.~N. Boddeti,
``AutoFHE: Automated adaption of CNNs for efficient evaluation over FHE,''
in \emph{Proc. USENIX Security}, 2024, pp.~2173--2190.

\bibitem{goodfellow2015adversarial}
I.~J. Goodfellow, J.~Shlens, and C.~Szegedy,
``Explaining and harnessing adversarial examples,''
in \emph{Proc. ICLR}, 2015.

\bibitem{cortes1995svm}
C.~Cortes and V.~Vapnik,
``Support-vector networks,''
\emph{Machine Learning}, vol.~20, no.~3, pp.~273--297, 1995.

\bibitem{bennett2000svm}
K.~P. Bennett and E.~J. Bredensteiner,
``Duality and geometry in SVM classifiers,''
in \emph{Proc. ICML}, 2000, pp.~57--64.

\bibitem{tolstikhin2021mlpmixer}
I.~O. Tolstikhin \emph{et al.},
``MLP-Mixer: An all-MLP architecture for vision,''
in \emph{Proc. NeurIPS}, 2021.

\bibitem{touvron2023resmlp}
H.~Touvron \emph{et al.},
``ResMLP: Feedforward networks for image classification with data-efficient training,''
\emph{IEEE TPAMI}, vol.~45, no.~4, pp.~5314--5321, 2023.

\bibitem{zhang2022glnn}
S.~Zhang, Y.~Liu, Y.~Sun, and N.~Shah,
``Graph-less neural networks: Teaching old MLPs new tricks via distillation,''
in \emph{Proc. ICLR}, 2022.

\bibitem{oquab2023dinov2}
M.~Oquab \emph{et al.}, ``DINOv2: Learning robust visual features without supervision,'' \emph{Trans. Mach. Learn. Res. (TMLR)}, 2024.

\bibitem{qwen2025qwen3embedding}
Y.~Zhang, M.~Li, D.~Long, X.~Zhang, H.~Lin, B.~Yang, P.~Xie, A.~Yang,
D.~Liu, J.~Lin, F.~Huang, and J.~Zhou,
``{Qwen3} Embedding: Advancing text embedding and reranking through foundation
models,''
arXiv:2506.05176v3, 2025.

\end{thebibliography}

\end{document}